\documentclass[usenatbib]{mn2e}
\usepackage{graphicx}
\usepackage{amsmath}

\def\apj{\rm ApJ}
\def\apjl{\rm ApJL}
\def\apjs{\rm ApJS}
\def\aj{\rm AJ}
\def\mnras{\rm MNRAS}
\def\nat{\rm Nature}
\def\pasj{\rm PASJ}
\def\pasp{\rm PASP}
\def\pasa{\rm PASA}
\def\aap{\rm AAP}
\def\araa{\rm ARA\&A}

\def\gax{\mathrel{\raise.3ex\hbox{$>$}\mkern-14mu\lower0.6ex\hbox{$\sim$}}}
\def\lax{\mathrel{\raise.3ex\hbox{$<$}\mkern-14mu\lower0.6ex\hbox{$\sim$}}}
\def\gtorder{\mathrel{\raise.3ex\hbox{$>$}\mkern-14mu
             \lower0.6ex\hbox{$\sim$}}}
\def\ltorder{\mathrel{\raise.3ex\hbox{$<$}\mkern-14mu
             \lower0.6ex\hbox{$\sim$}}}

\begin{document}

\title [Supernovae Producing Unbound Binaries and Triples ]
   {Supernovae Producing Unbound Binaries and Triples}

\author[C.~S. Kochanek]{ 
    C.~S. Kochanek$^{1,2}$
    \\
  $^{1}$ Department of Astronomy, The Ohio State University, 140 West 18th Avenue, Columbus OH 43210 \\
  $^{2}$ Center for Cosmology and AstroParticle Physics, The Ohio State University,
    191 W. Woodruff Avenue, Columbus OH 43210 \\
   }

\maketitle

\begin{abstract}
The fraction of stars which are in binaries or triples 
at the time of stellar death and the fraction of these systems which survive
the supernova (SN) explosion are crucial constraints for evolution models
and predictions for gravitational wave source populations.  These 
fractions are also subject to direct observational determination.
Here we search 10 supernova remnants (SNR) containing compact objects with proper
motions for unbound binaries or triples using Gaia EDR3
and new statistical methods and tests for false positives.  
We confirm the one known example of an unbound binary, HD~37424 in
G180.0$-$01.7, and find no other examples. Combining this with our
previous searches for bound and unbound binaries, and assuming no 
bias in favor of finding interacting binaries, we find that
72.0\% (52.2\%-86.4\%, 90\% confidence) of SN producing neutron stars are not
binaries at the time of explosion, 13.9\% (5.4\%-27.2\%) produce bound binaries
and 12.5\% (2.8\%-31.3\%) produce unbound binaries.  With a strong bias
in favor of finding interacting binaries, the medians shift to
76.0\% were not binaries at death, 9.5\% leave bound and 13.2\% leave unbound
binaries. Of explosions that do
not leave binaries, $<18.9\%$ can be fully unbound triples.  These limits
are conservatively for $M>5M_\odot$ stars, although the mass limits for
individual systems are significantly stronger. At birth, the progenitor
of PSR~J0538$+$2817 was probably a $13$-$19M_\odot$ star, and at the 
time of explosion it was probably a Roche limited, partially stripped star 
transferring mass to HD~37424 and then producing a Type~IIL or IIb 
supernova. 
\end{abstract}


\section{Introduction}

Massive stars are not isolated.  Almost all start in
binaries (e.g., \citealt{Kobulnicky2007}, 
\citealt{Sana2012}, \citealt{Moe2016}, \citealt{Villasenor2021}) and many are in triple or still higher 
order systems (\citealt{Moe2016}). The evolution of these systems is
crucial to understanding the expected properties of
supernovae (SN) and compact objects and there is a long
history of using binary population synthesis (BPS) models
to try and follow the evolution of binary populations
(e.g., \citealt{Belczynski2008}, \citealt{Eldridge2017}).
These population studies have recently expanded to follow the evolution
of triple or quadruple systems with their far richer
dynamical properties (e.g., \citealt{Toonen2020}, 
\citealt{Hamers2021}).  The end results for the 
evolution of massive star binaries and triples 
are particularly crucial to understanding the growing sample
of compact object mergers found through 
gravitational wave emission (e.g., \citealt{Abbott2021}).

While studies like \cite{Kobulnicky2007},
\cite{Sana2012}, \cite{Moe2016} and \cite{Villasenor2021}
 are excellent surveys of the multiplicity
of massive stars, the path to mergers or
interacting compact objects is long and complex.
In addition to standard problems in stellar evolution,
there are the added problems of mass transfer,
common envelope evolution, neutron star (or black
hole) birth kicks (see, e.g., the review of BPS
models by \citealt{Han2020})
 and the nature of the mapping
between core collapse and outcome (supernova or
failed supernovae, neutrons star or black hole 
etc., see, e.g., \citealt{Pejcha2015}, \citealt{Kochanek2014}, \citealt{Sukhbold2016}).  
All of the physical effects are both uncertain
and important for any prediction, including for the
population of merging compact objects.

This process would be better constrained if we understood
the multiplicity of systems containing compact objects
and the fraction of initially multiple systems that survive
the death of a member star.  Compact object mergers,
a small minority of the initial systems
lying at the very end a complex evolutionary history, are
likely a very poor way to understand this problem.  
Similarly, interacting neutron star and black hole binaries are a small.
biased fraction of binaries containing compact objects.
Searches for non-interacting neutron star (other than pulsars) and black hole binaries are
difficult and only just starting (e.g., \citealt{Thompson2019},
\citealt{Jayasinghe2021}) and searches
for unbound compact objects only become fully practical
as part of the microlensing component of the
Roman Space Telescope (e.g., \citealt{Lam2020}).

There is, however, a second point in the evolution of
these systems where it is possible 
to make progress. Studies of supernovae remnants (SNRs) can be
used to determine the multiplicity of massive stars just 
before and after the death of a member.  This was originally
conceived of as a search for O/B stars associated with
SNRs, potentially with peculiar kinematics 
(\citealt{vandenBergh1980}, \citealt{Guseinov2005}). 
However, current models (e.g., \citealt{Renzo2019}),
and data (e.g., \citealt{Kochanek2019}) suggest that the
kinematics of these stars are not very distinctive.
Nonetheless, given an estimate of the explosion site 
based on the centroid or expansion of the SNR, one can
try to identify stars consistent with that location
at the estimated time of explosion. 

The easiest systems to examine are the very youngest
SNRs: the Crab, Cas~A and SN~1987A.  Because of their
youth, the region that must be searched even for
unbound companions is so small that it is trivial
to rule out bound or unbound companions down to
mass limits of $\sim M_\odot$ or less (\citealt{Kochanek2018}, also
see \citealt{Kerzendorf2019} and \citealt{Fraser2019}).
The lack of a companion in the Type~IIb
(\citealt{Krause2008}, \citealt{Rest2011})
SN Cas~A 
is particularly interesting 
because models for producing Type~IIb SNe typically
rely on mass transfer (e.g., \citealt{Benvenuto2013}).  The lack of a companion
to SN~1987A is consistent with models of the progenitor
as a merger remnant (e.g., \citealt{Podsiadlowski1992}).

Older SNRs with identified compact remnants and limited
extinction are easy to search for bound remnants simply
using the properties of the optical counterpart to the
neutron star.  \cite{Kochanek2019} identified 23 
Galactic SNRs where it was practical to carry out such a search, 
finding no non-interacting binaries and three previously
identified interacting binaries.  One of these systems,
SS~433, is a Roche accreting system where the compact
object is generally believed to be a black hole (see, e.g., \citealt{Hillwig2008}).  The
other two systems, HESS~J0632$+$057 (\citealt{Hinton2009}) and
1FGL~J1018.6$-$5856 (\citealt{Corbet2011}) are neutron star, wind 
accretion, high mass X-ray binaries (HMXBs).  \cite{Kochanek2019} speculated that
shock heating of the companion by the SN blast wave
might drive it to have stronger winds for a period
of time after the SN explosion, making it easier to
produce such wind accretion systems.  The lack of 
non-interacting binaries implied that fewer than 9.1\%
(90\% confidence) of SN lead to such systems with
a companion stellar mass $>3 M_\odot$.

Identifying unbound binaries or higher multiplicity 
systems in older SNRs is more challenging because the
star and the stellar remnant are no longer co-located. 
\cite{vandenBergh1980} appears to have carried out
the first search, looking for an excess of O and B
stars in 17 SNRs and finding none. \cite{Guseinov2005}
examine 48 SNRs for O and B stars based on their
magnitudes, colors and proper motions, producing a
list of candidates.  \cite{Dincel2015} identified the
most promising candidate to date, HD~37424 in G180.0$-$0.17
by back-propagating the proper motions of the star
and the pulsar to find that they intersect near the
center of the SNR.  \cite{Boubert2017} examined 10
SNRs using the Tycho-Gaia astrometric solution 
(TGAS, \citealt{Michalik2015}) proper motions to search for stars
which passed close to the centers of the SNRs on
reasonable time scales.  They identify four candidates,
one of which is HD~37424. \cite{Lux2021} did the same
for 12 SNRs using Gaia DR2 proper motions. \cite{Fraser2019} used Gaia
DR2 proper motions for both the stars and the compact
objects in the Crab, Cas~A and Vela SNRs, confirming
the \cite{Kochanek2018} result for the Crab and Cas~A
and only finding very low luminosity candidates in Vela. 

Because many SNRs are asymmetric and distorted, using 
an estimate of the center of the SNR as an estimate of
the explosion site is likely fraught
for many SNRs unless a generous minimum distance
is allowed to compensate for potential
systematic problems.   It is much safer to use the cases
where a proper motion has been measured for the neutron
star and we can simply search for intersections of the
paths of the star and neutron star at some reasonable time in the past and 
distance from the center.  Here we carry out this search
for 7 neutron stars and 3 interacting binaries associated
with SNRs that have sufficiently good proper motion
measurements to carry out a search.
We describe the sample and our methods in \S2.  In
particular, we describe methods to evaluate the probability
of false positives both as a general assessment for each
SNR and for particular stellar candidates.  In \S3 we
present the results for unbound binaries.  In \S4 we 
extend the search to fully unbound triples.  In \S5 we consider
the one candidate in more detail. We discuss the
results in \S6.

\begin{table*}
  \centering
  \caption{Remnants }
  \begin{tabular}{llr}
  \hline
  \multicolumn{1}{c}{SNR} &
  \multicolumn{1}{c}{Center} &
  \multicolumn{1}{c}{Radius (')} \\
  \hline
G039.7$-$02.0 &12:12:20 $+$04:55:00  &$ 42.4$ \\
G069.0$+$02.7 &19:54:50 $+$33:00:30  &$ 40.0$ \\
G109.1$-$01.0 &23:01:39 $+$58:53:00  &$ 16.5$ \\
G130.7$+$03.1 &02:05:34 $+$64:49:50  &$  4.2$ \\
G180.0$-$01.7 &05:40:01 $+$27:48:09  &$100.0$ \\
G205.5$+$00.5 &06:39:00 $+$06:30:00  &$110.0$ \\
G260.4$-$03.4 &08:22:28 $-$42:57:29  &$ 30.0$ \\
G263.9$-$03.3 &08:35:21 $-$45:10:35  &$249.0$ \\
G284.3$-$01.8 &10:18:15 $-$59:00:00  &$12.0$  \\ 
G296.5$+$10.0 &12:09:40 $-$52:25:00  &$ 38.0$ \\
  \hline
  \end{tabular}
  \label{tab:snr}
\end{table*}

\begin{table*}
  \centering
  \caption{Compact Object Proper Motions and Parallaxes}
  \begin{tabular}{lccccl}
  \hline
  \multicolumn{1}{c}{SNR} &
  \multicolumn{1}{c}{Position} &
  \multicolumn{1}{c}{PM RA} &
  \multicolumn{1}{c}{PM Dec)} &
  \multicolumn{1}{c}{$\varpi$} & 
  \multicolumn{1}{c}{Type} \\
   &
  \multicolumn{1}{c}{(J2000)} &
  \multicolumn{1}{c}{(mas/year)} &
  \multicolumn{1}{c}{(mas/year)} &
  \multicolumn{1}{c}{(mas)} \\
  \hline
G039.7$-$02.0 &19:11:49.565 $+$04:58:57.82 &$-3.027\pm0.024$  &$-4.777\pm 0.024$ &$0.118\pm 0.023$ &binary, SS~433 \\
G069.0$+$02.7 &19:52:58.206 $+$32:52:40.51 &$-28.8\pm0.9$     &$-14.7\pm0.9$     &$(0.75\pm0.25)$ &single\\ 
G109.1$-$01.0 &23:01:08.295 $+$58:52:44.45 &$-6.4\pm0.6$      &$-2.3\pm0.6$      &$(0.28\pm0.05$ &single\\ 
G130.7$+$03.1 &02:05:37.922 $+$64:49:41.33 &$-1.40\pm0.160$   &$0.540\pm0.575$   &($0.36\pm0.14)$ &single\\ 
G180.0$-$01.7 &05:38:25.057 $+$28:17:09.16 &$-23.559\pm0.085$ &$52.766\pm0.079$  &$0.72\pm0.12$ &single\\ 
G205.5$+$00.5 &06:32:59.257 $+$05:48:01.16 &$-0.026\pm0.020$  &$-0.428\pm0.016$  &$0.540\pm0.023$ &binary, MWC~148\\
G260.4$-$03.4 &08:21:57.274 $-$43:00:17.33 &$-74.2\pm7.7$     &$-30.3\pm6.2$     &$(0.75\pm0.25)$ &single\\ 
G263.9$-$03.3 &08:35:20.611 $-$45:10:34.88 &$-49.68\pm0.06$   &$29.90\pm0.10$    &$3.5\pm0.2$ &single\\ 
G284.3$-$01.8 &10:18:55.587 $-$58:56:45.98 &$-6.454\pm0.013$  &$2.256\pm0.010$   &$0.227\pm0.010$ &binary, 2MASS~J10185560$-$5856459 \\
G296.5$+$10.0 &12:10:00.913 $-$52:26:28.30 &$-12\pm5$         &$9\pm 8$          &$(0.50 \pm 0.25)$ &single\\  
  \hline
\multicolumn{5}{l} {
   Parallaxes in parentheses are based on rough distance estimates not direct measurements.}\\
  \end{tabular}
  \label{tab:ns}
\end{table*}

\begin{table*}
  \centering
  \caption{CMD Selection Limits}
  \begin{tabular}{lcccccccc}
  \hline
  \multicolumn{1}{c}{Mass} &
  \multicolumn{1}{c}{Color} &
  \multicolumn{1}{c}{Mag} &
  \multicolumn{1}{c}{Color)} &
  \multicolumn{1}{c}{Mag} &
  \multicolumn{1}{c}{Color} &
  \multicolumn{1}{c}{Mag} &
  \multicolumn{1}{c}{Color} &
  \multicolumn{1}{c}{Mag} \\
  \hline
  \multicolumn{1}{c}{$>1M_\odot$} &
  \multicolumn{1}{c}{$<1.5$} &
  \multicolumn{1}{c}{$\hphantom{-}5.5$} &
  &
  &
  \multicolumn{1}{c}{$1.5$-$2.0$} &
  \multicolumn{1}{c}{$-1-3(B_P-R_P-1)$} &
  \multicolumn{1}{c}{$>2$} &
  \multicolumn{1}{c}{$-4$} \\
  \multicolumn{1}{c}{$>2M_\odot$} &
  \multicolumn{1}{c}{$<0.7$} &
  \multicolumn{1}{c}{$\hphantom{-}2.0$} &
  \multicolumn{1}{c}{$0.7$-$1.1$} &
  \multicolumn{1}{c}{$\hphantom{-}3.0$} &
  \multicolumn{1}{c}{$1.1$-$2.0$} &
  \multicolumn{1}{c}{$-1-3(B_P-R_P-1)$} &
  \multicolumn{1}{c}{$>2$} &
  \multicolumn{1}{c}{$-4$} \\
  \multicolumn{1}{c}{$>3M_\odot$} &
  \multicolumn{1}{c}{$<0.7$} &
  \multicolumn{1}{c}{$\hphantom{-}1.0$} &
  \multicolumn{1}{c}{$0.7$-$1.1$} &
  \multicolumn{1}{c}{$\hphantom{-}2.0$} &
  \multicolumn{1}{c}{$1.1$-$2.0$} &
  \multicolumn{1}{c}{$-1-3(B_P-R_P-1)$} &
  \multicolumn{1}{c}{$>2$} &
  \multicolumn{1}{c}{$-4$} \\
  \multicolumn{1}{c}{$>4M_\odot$} &
  \multicolumn{1}{c}{$<0.0$} &
  \multicolumn{1}{c}{$\hphantom{-}0.5$} &
  \multicolumn{1}{c}{$0.0$-$1.3$} &
  \multicolumn{1}{c}{$-1.9$} &
  \multicolumn{1}{c}{$1.3$-$2.0$} &
  \multicolumn{1}{c}{$-1-3(B_P-R_P-1)$} &
  \multicolumn{1}{c}{$>2$} &
  \multicolumn{1}{c}{$-4$} \\
  \multicolumn{1}{c}{$>5M_\odot$} &
  \multicolumn{1}{c}{$<0.0$} &
  \multicolumn{1}{c}{$-0.4$} &
  \multicolumn{1}{c}{$0.0$-$1.3$} &
  \multicolumn{1}{c}{$-1.9$} &
  \multicolumn{1}{c}{$1.3$-$2.0$} &
  \multicolumn{1}{c}{$-1-3(B_P-R_P-1)$} &
  \multicolumn{1}{c}{$>2$} &
  \multicolumn{1}{c}{$-4$} \\
  \hline
\multicolumn{9}{l} {
  Mag gives the upper limit on $M_G$ for each $B_P-R_p$ color range. }
  \end{tabular}
  \label{tab:maglim}
\end{table*}

\begin{table*}
  \centering
  \caption{Searching For Disrupted Binaries}
  \begin{tabular}{lcrrrrrrrrrr}
  \hline
  \multicolumn{1}{c}{SNR} &
  \multicolumn{1}{c}{$M_{lim}$} &
  \multicolumn{1}{c}{$N_*$} &
  \multicolumn{3}{c}{$90\%$} &
  \multicolumn{3}{c}{$95\%$} &
  \multicolumn{3}{c}{$99\%$} \\
   &
  \multicolumn{1}{c}{($M_\odot$)}
   &
  &\multicolumn{1}{c}{$N_{obs}$}
  &\multicolumn{1}{c}{$N_{ran}$}
  &\multicolumn{1}{c}{Prob.}
  &\multicolumn{1}{c}{$N_{obs}$}
  &\multicolumn{1}{c}{$N_{ran}$}
  &\multicolumn{1}{c}{Prob.}
  &\multicolumn{1}{c}{$N_{obs}$}
  &\multicolumn{1}{c}{$N_{ran}$}
  &\multicolumn{1}{c}{Prob.} \\
  \hline
G039.7$-$02.0 &$>5$ &953 &  0 & 0.0 &   -- &  0 & 0.0 &   -- &  0 & 0.0 &   -- \\
            &$>4$ &957 &  0 & 0.0 &   -- &  0 & 0.0 &   -- &  0 & 0.0 &   -- \\
            &$>3$ &2202 &  0 & 0.0 &   -- &  0 & 0.0 &   -- &  0 & 0.0 &   -- \\
            &$>2$ &2527 &  0 & 0.0 &   -- &  0 & 0.0 &   -- &  0 & 0.0 &   -- \\
            &$>1$ &2641 &  0 & 0.0 &   -- &  0 & 0.0 &   -- &  0 & 0.0 &   -- \\
\hline
G069.0$+$02.7 &$>5$ &1382 &  0 & 0.1 &   -- &  0 & 0.2 &   -- &  2 & 0.7 & 16\% \\
            &$>4$ &1421 &  0 & 0.1 &   -- &  0 & 0.3 &   -- &  2 & 0.9 & 23\% \\
            &$>3$ &5178 &  2 & 1.9 & 55\% &  3 & 3.2 & 61\% & 20 &14.9 & 12\% \\
            &$>2$ &9158 &  9 & 7.8 & 38\% & 19 &15.1 & 19\% & 56 &46.6 & 10\% \\
            &$>1$ &23748 & 66 &83.7 & 98\% &111 &128.6 & 95\% &213 &227.5 & 84\% \\
\hline
G109.1$-$01.0 &$>5$ &26 &  0 & 0.1 &   -- &  0 & 0.1 &   -- &  0 & 0.2 &   -- \\
            &$>4$ &33 &  0 & 0.2 &   -- &  0 & 0.2 &   -- &  0 & 0.4 &   -- \\
            &$>3$ &243 &  1 & 1.2 & 71\% &  2 & 1.8 & 52\% &  3 & 2.8 & 53\% \\
            &$>2$ &519 &  1 & 1.2 & 71\% &  2 & 1.8 & 52\% &  3 & 2.8 & 53\% \\
            &$>1$ &1741 &  1 & 1.2 & 71\% &  2 & 1.8 & 52\% &  3 & 2.8 & 53\% \\
\hline
G130.7$+$03.1 &$>5$ &5 &  0 & 0.0 &   -- &  0 & 0.0 &   -- &  0 & 0.0 &   -- \\
            &$>4$ &5 &  0 & 0.0 &   -- &  0 & 0.0 &   -- &  0 & 0.0 &   -- \\
            &$>3$ &69 &  0 & 0.0 &   -- &  0 & 0.0 &   -- &  0 & 0.0 &   -- \\
            &$>2$ &155 &  0 & 0.0 &   -- &  0 & 0.0 &   -- &  0 & 0.0 &   -- \\
            &$>1$ &361 &  0 & 0.0 &   -- &  0 & 0.0 &   -- &  0 & 0.0 &   -- \\
\hline
G180.0$-$01.7 &$>5$ &0 &  1 & 0.0 &  0\% &  1 & 0.0 &  0\% &  1 & 0.0 &  3\% \\
            &$>4$ &0 &  1 & 0.0 &  2\% &  1 & 0.0 &  3\% &  1 & 0.1 &  7\% \\
            &$>3$ &0 &  1 & 0.2 & 18\% &  1 & 0.3 & 24\% &  1 & 0.5 & 39\% \\
            &$>2$ &29 &  2 & 0.8 & 18\% &  2 & 1.1 & 29\% &  2 & 1.9 & 55\% \\
            &$>1$ &8973 &  8 & 9.7 & 75\% &  9 &13.6 & 92\% & 18 &20.9 & 77\% \\
\hline
G205.5$+$00.5 &$>5$ &37 &  0 & 0.0 &   -- &  0 & 0.0 &   -- &  0 & 0.0 &   -- \\
            &$>4$ &54 &  0 & 0.0 &   -- &  0 & 0.0 &   -- &  0 & 0.0 &   -- \\
            &$>3$ &272 &  0 & 0.0 &   -- &  0 & 0.0 &   -- &  0 & 0.0 &   -- \\
            &$>2$ &1067 &  0 & 0.0 &   -- &  0 & 0.0 &   -- &  0 & 0.0 &   -- \\
            &$>1$ &11860 &  0 & 0.0 &   -- &  0 & 0.0 &   -- &  0 & 0.0 &   -- \\
\hline
G260.4$-$03.4 &$>5$ &147 &  0 & 0.2 &   -- &  0 & 0.3 &   -- &  0 & 0.7 &   -- \\
            &$>4$ &168 &  0 & 0.6 &   -- &  0 & 0.7 &   -- &  0 & 1.2 &   -- \\
            &$>3$ &658 &  0 & 1.9 &   -- &  1 & 2.8 & 94\% &  8 & 5.6 & 20\% \\
            &$>2$ &1079 &  4 & 5.7 & 82\% &  6 & 7.8 & 79\% & 14 &12.2 & 34\% \\
            &$>1$ &2679 & 28 &31.8 & 77\% & 35 &39.4 & 78\% & 54 &54.4 & 54\% \\
\hline
G263.9$-$03.3 &$>5$ &5 &  0 & 0.0 &   -- &  0 & 0.0 &   -- &  0 & 0.0 &   -- \\
            &$>4$ &16 &  0 & 0.0 &   -- &  0 & 0.0 &   -- &  0 & 0.0 &   -- \\
            &$>3$ &45 &  0 & 0.0 &   -- &  0 & 0.0 &   -- &  0 & 0.0 &   -- \\
            &$>2$ &97 &  0 & 0.0 &   -- &  0 & 0.0 &   -- &  0 & 0.0 &   -- \\
            &$>1$ &963 &  0 & 0.2 &   -- &  0 & 0.3 &   -- &  0 & 0.4 &   -- \\
\hline
G284.3$-$01.8 &$>5$ &322 &  0 & 0.0 &   -- &  0 & 0.0 &   -- &  0 & 0.0 &   -- \\
            &$>4$ &539 &  0 & 0.0 &   -- &  0 & 0.0 &   -- &  0 & 0.0 &   -- \\
            &$>3$ &1394 &  0 & 0.0 &   -- &  0 & 0.0 &   -- &  0 & 0.0 &   -- \\
            &$>2$ &2016 &  0 & 0.0 &   -- &  0 & 0.0 &   -- &  0 & 0.0 &   -- \\
            &$>1$ &2439 &  0 & 0.0 &   -- &  0 & 0.0 &   -- &  0 & 0.0 &   -- \\
\hline
G296.5$+$10.0 &$>5$ &63 &  0 & 0.0 &   -- &  0 & 0.0 &   -- &  0 & 0.0 &   -- \\
            &$>4$ &63 &  0 & 0.0 &   -- &  0 & 0.0 &   -- &  0 & 0.0 &   -- \\
            &$>3$ &606 &  0 & 0.2 &   -- &  0 & 0.3 &   -- &  0 & 0.3 &   -- \\
            &$>2$ &1305 &  2 & 0.8 & 21\% &  2 & 1.1 & 30\% &  2 & 1.3 & 37\% \\
            &$>1$ &8372 & 15 &13.9 & 42\% & 18 &15.8 & 33\% & 22 &17.5 & 17\% \\
\hline
  \hline
  \end{tabular}
  \label{tab:searchbin}
\end{table*}

\begin{table*}
  \centering
  \caption{Disrupted Binary Candidates}
  \begin{tabular}{lrcrrrrrrl}
  \hline
  \multicolumn{1}{c}{SNR} &
  \multicolumn{1}{c}{Gaia EDR3 ID}   &
  \multicolumn{1}{c}{$M_{lim}$} &
  \multicolumn{1}{c}{False}     &
  \multicolumn{1}{c}{$\chi_\varpi^2$} &
  \multicolumn{1}{c}{$\chi_\mu^2$} &
  \multicolumn{1}{c}{$t_{min}$} &
  \multicolumn{1}{c}{$R/R_{SNR}$} & 
  \multicolumn{1}{c}{$\varpi$}  &
  \multicolumn{1}{c}{Comment} \\
  \multicolumn{1}{c}{} &
  \multicolumn{1}{c}{} &
  \multicolumn{1}{c}{($M_\odot$)} &
  \multicolumn{1}{c}{(\%)} &
  \multicolumn{1}{c}{} &
  \multicolumn{1}{c}{} &
  \multicolumn{1}{c}{($10^3$~yr)} &
  \multicolumn{1}{c}{} &
  \multicolumn{1}{c}{(mas)} \\
  \hline
G069.0$+$02.7 &  2034331673776151552 &5 & 49 &$ 3.858$ &$ 0.062$ &$-57.7$ &$0.19$ &$  0.259 \pm  0.010 $ & \\ 
              &  2034320300712461184 &5 & 54 &$ 4.151$ &$ 0.461$ &$-36.7$ &$0.15$ &$  0.242 \pm  0.017 $ & \\ 
\hline 
G109.1$-$01.0 &  2013340789887658240 &3 &  4 &$ 3.073$ &$ 0.368$ &$-53.0$ &$0.15$ &$  0.365 \pm  0.012 $ & \\ 
              &  2013340381858998016 &3 & 64 &$ 3.479$ &$ 1.284$ &$-53.6$ &$0.15$ &$  0.210 \pm  0.033 $ & \\ 
              &  2013340175700592384 &3 & 41 &$ 1.220$ &$ 1.178$ &$-56.6$ &$0.17$ &$  0.330 \pm  0.022 $ & \\ 
\hline 
G180.0$-$01.7 &  3441732292729818752 &5 &  0 &$ 0.282$ &$ 0.261$ &$-29.3$ &$0.10$ &$  0.658 \pm  0.028 $ &HD37424 \\ 
              &  3441706518633166080 &2 & 14 &$ 0.433$ &$ 0.048$ &$-48.7$ &$0.14$ &$  0.798 \pm  0.018 $ & \\ 
\hline 
G260.4$-$03.4 &  5526330988095223552 &3 &100 &$ 3.890$ &$ 0.501$ &$-7.3$ &$0.12$ &$  0.268 \pm  0.052 $ & \\ 
              &  5526330472698907904 &3 &100 &$ 5.723$ &$ 0.026$ &$-5.5$ &$0.04$ &$  0.167 \pm  0.055 $ & \\ 
              &  5526330403979671552 &3 &100 &$ 6.161$ &$ 0.343$ &$-7.0$ &$0.11$ &$  0.156 \pm  0.072 $ & \\ 
              &  5526325387457616128 &3 &100 &$ 3.051$ &$ 0.310$ &$-4.0$ &$0.03$ &$  0.321 \pm  0.046 $ & \\ 
              &  5526325284378390272 &3 & 99 &$ 5.250$ &$ 0.006$ &$-2.0$ &$0.12$ &$  0.181 \pm  0.029 $ & \\ 
              &  5526325284378390144 &3 & 99 &$ 3.631$ &$ 0.309$ &$-2.4$ &$0.10$ &$  0.280 \pm  0.042 $ & \\ 
              &  5526319099625741184 &3 & 78 &$ 4.923$ &$ 0.007$ &$-9.0$ &$0.19$ &$  0.208 \pm  0.054 $ & \\ 
              &  5526319095324073856 &3 & 94 &$ 5.814$ &$ 0.011$ &$-8.7$ &$0.18$ &$  0.189 \pm  0.091 $ & \\ 
\hline 
G296.5$+$10.0 &  6125329220509401984 &2 & 41 &$ 0.011$ &$ 1.480$ &$-10.2$ &$0.16$ &$  0.474 \pm  0.030 $ & \\ 
              &  6125328945636431104 &2 & 76 &$ 2.250$ &$ 0.128$ &$-14.8$ &$0.19$ &$  0.152 \pm  0.093 $ & \\ 
\hline 
  \hline
  \end{tabular}
  \label{tab:candbin}
\end{table*}

\begin{table*}
  \centering
  \caption{Disrupted Triple Candidates}
  \begin{tabular}{llllrrrrrl}
  \hline
  \multicolumn{1}{c}{SNR} &
  \multicolumn{1}{c}{Gaia EDR3 ID}   &
  \multicolumn{1}{c}{$M_{lim}$} &
  \multicolumn{1}{c}{False}     &
  \multicolumn{1}{c}{$\chi_\varpi^2$} &
  \multicolumn{1}{c}{$\chi_\mu^2$} &
  \multicolumn{1}{c}{$t_{min}$} &
  \multicolumn{1}{c}{$R/R_{SNR}$} &
  \multicolumn{1}{c}{$\varpi$}  &
  \multicolumn{1}{c}{Comment} \\
  \multicolumn{1}{c}{} &
  \multicolumn{1}{c}{} &
  \multicolumn{1}{c}{($M_\odot$)} &
  \multicolumn{1}{c}{(\%)} &
  \multicolumn{1}{c}{} &
  \multicolumn{1}{c}{} &
  \multicolumn{1}{c}{($10^3$~yr)} &
  \multicolumn{1}{c}{} &
  \multicolumn{1}{c}{(mas)} \\
  \hline
G069.0$+$02.7 &  2034319682237156096,   2034319510438472320 &3,3 & 66, 68 &$ 5.689$ &$ 1.047$ &$-42.6$ &$0.10$ &$  0.197 \pm  0.018 $ & \\ 
              &  2034331570696943232,   2034330952221647232 &3,2 & 65, 30 &$ 4.817$ &$ 3.804$ &$-52.6$ &$0.13$ &$  0.271 \pm  0.021 $ & \\ 
              &  2034342943771429504,   2034329848377469824 &3,1 & 72,  1 &$ 3.994$ &$ 1.831$ &$-53.9$ &$0.15$ &$  0.344 \pm  0.022 $ & \\ 
              &  2034342943771428992,   2034331227099572480 &3,1 & 56, 65 &$ 3.303$ &$ 3.225$ &$-49.4$ &$0.11$ &$  0.447 \pm  0.012 $ & \\ 
              &  2034342943771428992,   2034319299937887872 &3,1 & 56, 42 &$ 1.804$ &$ 7.166$ &$-47.4$ &$0.10$ &$  0.436 \pm  0.014 $ & \\ 
\hline 
G180.0$-$01.7 &  3441732292729818752,   3442478238354719872 &5,1 &  0, 35 &$ 0.283$ &$ 1.769$ &$-29.3$ &$0.10$ &$  0.658 \pm  0.026 $ &HD37424 \\ 
\hline 
G260.4$-$03.4 &  5526325353097874176,   5526325250018655616 &2,2 & 99, 98 &$ 1.189$ &$ 6.940$ &$-3.2$ &$0.07$ &$  0.667 \pm  0.008 $ & \\ 
\hline 
G296.5$+$10.0 &  6125329220509401984,   6125328911271728768 &2,1 & 41, 92 &$ 2.128$ &$ 4.753$ &$-10.0$ &$0.16$ &$  0.496 \pm  0.026 $ & \\ 
\hline 
  \end{tabular}
  \label{tab:candtrip}
\end{table*}

\section{Searching for Unbound Systems}

The 10 SNRs we examine are summarized in
Appendix A, and their properties are given in Tables~\ref{tab:snr}
and \ref{tab:ns}.   Only two of the single neutron stars have directly
measured parallaxes.  For the remaining five we use rough distance
estimates for the SNR expressed as parallaxes.  For these cases the parallax
entry in Table~\ref{tab:ns} is enclosed by parentheses. The
three interacting binaries have Gaia proper motions and parallaxes.

We select nearby stars from Gaia EDR3 (\citealt{Gaia2016}, \citealt{Gaia2021}), requiring them to
have proper motions, all three Gaia magnitudes 
($G$, $B_P$ and $R_P$) and $G<18$~mag.  We include no restrictions
on the RUWE statistic for the quality of the parallax, and
for the present effort we can ignore the small systematic
uncertainties in the parallax zero point and the statistical
differences between the inverse parallax and the distance.
We are assuming that any unbound star has moved away from
the compact object, which is a very safe assumption since
even a 10~km/s velocity difference produces an angular 
separation of 2~arcsec at 10~kpc after $10^4$ years. 

The bright $G \simeq 3$~magnitude limit of Gaia could be a problem
for nearby SNRs because it may lead to the exclusion of the most
luminous stars.  We addressed this by also searching the Hipparcos 
(\citealt{Perryman1997}) and Bright Star (\citealt{Hoffleit1995})
catalogs. For Hipparcos, we used the updated astrometric solution 
from \cite{vanLeeuwen2007}.  For the search regions used here,
no bright stars were missing from the Gaia catalog.  

We used PARSEC (\citealt{Pastorelli2020}) isochrones to group the stars into mass ranges based on
their locations in an extinction-corrected $M_G$ versus $C=B_P-R_P$ color
magnitude diagram. This is the only part of the calculation which uses
distances, and simply using the inverse parallax here is sufficiently
accurate because of the steepness of stellar mass-luminosity
relations.  We grouped them using limits of $1$, $2$, $3$, $4$ and $5 M_\odot$,
with the various color-dependent absolute magnitude limits given in 
Table~\ref{tab:maglim}.  As we noted in \cite{Kochanek2019}, the lower mass
stars will not yet be on the main sequence when the massive
stars begin to explode as supernovae.  These pre-main sequence
stars lie almost a full magnitude redwards in color from the
zero age main sequence.  The jump downwards in the magnitude limit
near $C=B_P-R_P=1$ is to ensure the inclusion of these 
pre-main sequence stars.  For higher mass stars, the ``dip'' to fainter, 
redder pre-main sequence stars is no longer necessary. For the
$>1M_\odot$ case, we also exclude stars with 
$-1-3(C-1) < M_G < 3+5(C-1)$ for $1.1 < C < 1.5$ to cut out
the lower part of the red giant branch.   The $N_*$ column of
Table~\ref{tab:searchbin} gives the number of stars above each
$M_{lim}$ that could have intersected the path of the neutron
star in the last $10^5$~years at a velocity below $400$~km/s
(see below).  Magnitude limits are becoming important when the
number of stars stops rapidly increasing with $M_{lim}$.

For extinction estimates we used the three dimensional {\tt combined19 mwdust}
models (\citealt{Bovy2016}), which combine the 
\cite{Drimmel2003}, \cite{Marshall2006}
and \cite{Green2019b} models to provide extinction estimates for any sky position
as a function of distance. We
extracted the V band extinction, and use the PARSEC estimates of 
$A_\lambda/A_V$ for the Gaia EDR3 bands and an $R_V=3.1$ extinction law 
to convert the V band extinction to those for the $G$, $B_P$ and $R_P$ bands.

We first simply identify stars or pairs of stars whose parallaxes
are consistent with the parallax of the neutron star.
Given parallax estimates of $\varpi_i \pm \sigma_{\varpi,i}$
for $i=1,2$ (binary) or $i=1,2,3$ (triple) we adopt
the joint best-fit parallax of
\begin{equation}
   \varpi = \left[ \sum_i { \varpi_i \over \sigma_{\varpi,i}^2 } \right]
         \left[ \sum_i { 1 \over \sigma_{\varpi,i}^2 } \right]^{-1}
\end{equation}
and the goodness of fit for this best-fit parallax,
\begin{equation}
    \chi^2_\varpi = \sum_i \left( { \varpi_i - \varpi \over \sigma_{\varpi,i}}\right)^2
    \label{eqn:chipar}.
\end{equation}
We initially keep all stars which as a potential binary companion
satisfy $\chi^2_\varpi < 9$.  For the neutron stars without direct parallaxes,
we use the weak constraints based on estimates of the distance to
the SN discussed in the Appendix.  

Next we restricted the sample to stars which could intersect the path
of the neutron star in the time interval $-t_m < t < 0$ at a stellar velocity less
than $v_* < v_{max}$.  We again use a generous set of criteria because
we are not yet using the actual stellar proper motions and their uncertainties,
but this step does significantly reduce the numbers of stars which must
be considered.  We set $t_m = 10^5$~years and $v_{max} = 400$~km/s.
If for distance $d$ the position difference between the neutron star and the star is
$\vec{\Delta x} = d \vec{\Delta\alpha}$ where
$\vec{\Delta\alpha} = \vec{\alpha}_{ns}-\vec{\alpha}_*$, and
the velocity of the neutron star is $\vec{v}_{ns} = d\vec{\mu}_{ns}$,
then the stellar velocity $v_*$
required to intercept the compact object at time $t$ is
\begin{equation}
   v_*^2(t) = v_{ns}^2 + 2 t^{-1} \vec{v}_{ns} \cdot \vec{\Delta x} + \Delta x^2 t^{-2}
\end{equation}
which has a minimum of
\begin{equation}
   v_{*min}^2 = { \left| \vec{\Delta x} \times \vec{v}_{ns}\right| ^2 \over \Delta x^2 }
\end{equation}
at time
\begin{equation}
   t_{*min}^{-1} = -{ \vec{\Delta x} \cdot \vec{v}_{ns} \over \Delta x^2 }.
\end{equation}
If the time of minimum velocity does not lie in the interval
$-t_m < t_{*min} < 0$, then the minimum velocity for the star
to intercept the compact object is just $v_*(-t_m)$, while if
it does lie in the interval, the minimum velocity needed is
$v_{*min}$.  The angular coordinate differences include the
necessary $\cos \delta $ term for the separation in RA, and the 
positions of the stars passing the criteria are close enough to 
the neutron star that we need not do a full spherical trigonometry 
calculation.  

We do not want to restrict our search to stars which can exactly
intersect the neutron star at a particular time.  If we allow binary
separations of $a_{max}$, then a star need only pass within the angular 
distance
\begin{equation} 
   \sigma_a = 10\farcs0 \left( { a_{max} \over 10^4~\hbox{AU} } \right)
        \left( { \varpi \over \hbox{mas}} \right)
    \label{eqn:matchdist}
\end{equation}
to potentially have been a binary companion of the neutron star. 
So we need a method that accounts for both the uncertainties
in the proper motions and a finite angular separation 
at the time of explosion.  We can account for both by introducing
the true proper motions of the objects as an intermediate variable.
Consider the motion in RA.  Let the position separation be 
$\Delta\alpha_{12}=\alpha_1-\alpha_2$ (the $\cos \delta$ is
again implicit),
the measured proper motions in RA be $\mu_{\alpha,i}$, and 
the true proper motions in RA be  $\mu_{\alpha,i}^T$.  The
true proper motion is constrained by the measured proper
motions, and the separation as a function of time then 
depends on $\Delta\alpha_{12}-t(\mu_{\alpha,1}^T-\mu_{\alpha,2}^T)$.
This leads to the fit statistic
\begin{eqnarray}
\chi^2_{\alpha,12} &= 
 { \left( \mu_{\alpha,1}-\mu_{\alpha,1}^T\right)^2 \over \sigma_{\alpha,1}^2 }
+{ \left( \mu_{\alpha,2}-\mu_{\alpha,2}^T\right)^2 \over \sigma_{\alpha,2}^2 } \\
&+{ \left( \Delta\alpha_{12} - t(\mu_{\alpha,1}^T-\mu_{\alpha,2}^T)\right)^2 \over
    \sigma_a^2 } \nonumber
\end{eqnarray}
where the first two terms constrain the true proper motions using the
measured proper motions and the last term constrains the closest approach
given the true proper motions.  If we optimize the statistic with respect
to the unknown true proper motions,   
\begin{equation}
\chi^2_{\alpha,12} = 
{ \left( \Delta\alpha_{12}/t - \Delta\mu_{\alpha,12}\right)^2 \over
    \sigma_{\alpha,12}^2 }
\end{equation}
where $\Delta\mu_{\alpha,12}=\mu_{\alpha,1}-\mu_{\alpha,2}$, and
\begin{equation}
  \sigma_{\alpha,12}^2 = \sigma_{\alpha,1}^2 + \sigma_{\alpha,2}^2 +\sigma_a^2/t^2.
   \label{eqn:error1}
\end{equation}
If we now define $\Delta_{\alpha,12}=\Delta\alpha_{12}/t-\Delta\mu_{\alpha,12}$, then
the overall statistic is
\begin{equation}
  \chi^2_\mu = 
  { \Delta_{\alpha,12}^2 \over \sigma_{\alpha,12}^2 } +
  { \Delta_{\delta,12}^2 \over \sigma_{\delta,12}^2 },
   \label{eqn:fit}
\end{equation}
after including the analogous term in Declination.  In essence, a $\chi^2_{\mu} \simeq 3$
means that the star can pass within distance $a_{max}$ of the neutron star using proper motions that
are each within $1\sigma$ of the measured values.  While the optimization of $\chi^2_{\mu}$ is formally
non-linear in $t^{-1}$ due to Eqn.~\ref{eqn:error1}, the problem is easily solved iteratively:
fix the time in Eqn.~\ref{eqn:error1} to determine a new optimal time, use this 
time to update Eqn.~\ref{eqn:error1} and repeat.  The ratio $ t \sigma_\mu / \sigma_a$ 
of the two terms in Eqn.~\ref{eqn:error1},
indicates whether the position uncertainty due 
to the proper motions at closest approach ($t_m\sigma_\mu $) is larger ($t_m\sigma_\mu/\sigma_a > 1$) or smaller 
($t_m\sigma_\mu/\sigma_a<1$) than the desired maximum separation $\sigma_a$.  
While we do not include the position uncertainties because they are unimportant,
they can be included as an error term like $\sigma_a$.

With the uncertainties fixed, the closest approach time is
\begin{equation}
    t_m = - { \Delta\alpha_{12}^2\sigma_{\delta,12}^2 + \Delta\delta_{12}^2\sigma_{\alpha,12}^2 \over
       \Delta\alpha_{12}\Delta\mu_{\alpha,12}\sigma_{\delta,12}^2+
       \Delta\delta_{12}\Delta\mu_{\delta,12}\sigma_{\alpha,12}^2}
     \label{eqn:time}
\end{equation}
with a goodness of fit
\begin{equation}
   \chi^2_{12} = 
   { \left( \Delta\delta_{12} \Delta\mu_{\alpha,12} - \Delta\alpha_{12} \Delta\mu_{\delta,12} \right)^2
     \over \Delta\alpha_{12}^2 \sigma_{\delta,12}^2  + \Delta\delta_{12}^2 \sigma_{\alpha,12}^2 }.
    \label{eqn:binary}
\end{equation}
For equal uncertainties, the goodness of fit is basically just the amplitude
of the proper motions perpendicular to the separation vector relative to the proper
motion uncertainties, squared.  

We use the same procedure to search for triples.  The overall fit statistic 
replacing the first term of Eqn.~\ref{eqn:fit} is
\begin{equation}
 \chi^2_\alpha =
  { \sigma_{\alpha,3}^2 \Delta_{\alpha,12}^2 + 
    \sigma_{\alpha,2}^2 \Delta_{\alpha,13}^2 +
    \sigma_{\alpha,1}^2 \Delta_{\alpha,23}^2 \over
  \sigma_{\alpha,1}^2 \sigma_{\alpha,2}^2 + 
  \sigma_{\alpha,1}^2 \sigma_{\alpha,3}^2 + 
  \sigma_{\alpha,2}^2 \sigma_{\alpha,3}^2 }
    \label{eqn:triple}
\end{equation}
with an equivalent term for the Declination. The
definitions of the errors $\sigma_{\alpha,i}$ depend on how we want to restrict 
the separations.  If we want to only restrict the
minimum distance of the stars from the neutron star, then 
the uncertainties are $\sigma_{\alpha,1}$ for
the neutron star and 
$\sigma_{\alpha,i}^2 \rightarrow \sigma_{\alpha,i}^2+\sigma_a^2/t^2$ for 
each of the two stars. Eqn.~\ref{eqn:triple} then becomes
identical to Eqn.~\ref{eqn:fit} in the limit of
making the proper motion uncertainties of one of the stars
infinite.  If you also add a term constraining the separation
of the two stars by $\sigma_a$, then the uncertainties are
$\sigma_{\alpha,i}^2 \rightarrow \sigma_{\alpha,i}^2+\sigma_a^2/3 t^2$
for all three objects.  Here we use this latter form.
The solution is again found by
iteratively holding the uncertainties fixed, solving for $t_m$,
and updating the uncertainties.  

We consider the problem of false positives in two ways.  First, we
generally characterize each SNR by scrambling the stellar positions
relative to their magnitudes, proper motions and parallaxes.  We keep the
latter three quantities tied because the uncertainties in the proper 
motions and parallaxes are in large part determined by the magnitude. While
this is also true of the uncertainties in the positions, the position 
uncertainties are unimportant compared to the proper motion uncertainties
and can be regarded as fixed.  For each trial we assigned a random number 
to each star and assigned positions in the order of these random numbers.
So if star \#1 appeared as the fiftieth entry in this list, it was assigned
the position of star \#50.  We rejected the occasional cases where a
star would be assigned its true position, but we did not try to eliminate
the occasional cases where multiple trials would consider the same random
pairing.  With $N_* \simeq 4000$ stars and $N_t=100$ trials, there are
$N_* N_t \simeq 400,000$ random pairings of the neutron star with single stars
of which we expect only $ \sim 100$ duplicates.  
We do $N_t=100$ random trials for each
SNR and count the number of candidates which have a position of
closest approach inside 20\% of the SNR radius and $\chi^2_\varpi+\chi^2_\mu$
smaller than several threshold values.   

We can also evaluate the likelihood that a particular candidate is a
false positive.  This can be very different than the general probability
of a false positive if a candidate has an unusual proper motion.  We 
take all stars with $\chi^2_\varpi < 9$ and within 1~mag of each 
candidate unbound binary star, and put it at the position of the
candidate.  We restrict the magnitude range so that the stars have
similar kinematics and proper motion uncertainties.  We then determine the fraction
of these trials which would produce a $\chi^2_\varpi + \chi^2_\mu$
goodness of fit less than the larger of the fit statistic of
the candidate or 2.706 (90\% confidence).  We only do this for the
unbound binary candidates because all candidate unbound triples are
pairs of binary candidates.  

\section{Binaries}
\label{sec:binaries}

Table~\ref{tab:searchbin} summarizes the overall statistics of 
searching for unbound stars in each of the SNRs.  For probability
limits of 90\% ($\chi^2_\varpi+\chi^2_\mu < 2.706$), 95\% ($3.841$)
and 99\% ($6.635$), the table gives the number of stars considered,
$N_*$, the number of actual candidates,
$N_{obs}$, the average number of random candidates found per trial,
$N_{ran}$, and the Poisson likelihood of finding $N_{obs}$ or more
candidates given an expectation value of $N_{ran}$, all as function 
of the minimum mass selection limit $M_{lim}$.  
The probability is meaningless for a bin with $N_{obs}=0$,
so no probability is given for these bins.  
These are all for 
$R/R_{SNR}<0.2$, $a_{max}=10^4$~AU and $t_{min}>-10^5$~years.  
Note that there is little benefit from limiting the search to more compact binaries
(smaller $a_{max}$) because this scale does not matter once it is smaller
than the position uncertainties created by back propagating the 
proper motions for tens of thousands of years.

The SNRs span a range of statistical
properties from having essentially zero chance of a false positive
for G263.9$-$03.3 because of its proximity and accurate neutron star parallax,
to expecting large numbers of false positives for G069.0$+$02.7
because of the large number of stars and the absence of a neutron star 
parallax.  The expected number of false positives rises as the 
mass limit is lowered because both the number of stars and the
uncertainties in the Gaia parallaxes and proper motions increase,
and it increases as the probability threshold is weakened.
The numbers of observed candidates tracks the expected number from
the random realizations extremely well, so the Poisson probabilities
of finding at least as many candidates as observed are almost always
significant.  
In fact, the only SNR
with a candidate that is very unlikely to be a false positive is
G180.0$-$01.7, and this is the \cite{Dincel2015} candidate
HD~37424. 

Since we are primarily interested in the more massive companions,
we examined the candidates in the highest mass limit with candidates
for each SNR, with the exception of G180.0$-$01.7, where use the
mass limit producing a second candidate besides HD~37424.  The
properties of these candidates are given in Table~\ref{tab:candbin}.
SNRs with no candidates are not included.
In Table~\ref{tab:candbin} we include the probability of obtaining
a $\chi^2_\varpi+\chi^2_\mu$ value less than the larger of $2.706$
and the actual value
when assigning the same position the parallaxes and proper motions
of some other star within $\pm 1$~G~mag of the candidate.  
Essentially all the candidates except HD~37424 have positions 
that produce equally good or better fit statistics for significant 
fractions of randomly selected proper motions.

In the end, we reject all the candidates in Table~\ref{tab:candbin}
except HD~37424.  The two candidates in G069.0$+$02.7 are likely false
positives --  their distances imply they are well behind the SNR, and the
intercept times are somewhat young.  For G109.1$-$01.0, one candidate
does have a low false positive probability (4\%), but the parallaxes
are all somewhat high or low and the intercept times are too old.
Keep in mind that the false positive rates should be interpreted 
with a ``trials'' factor in mind -- we are selecting multiple stars
for having a path intersecting the neutron star from a still larger sample, so
the probability of a 4\% false positive probability star being a false
positive is larger than 4\% because of the multiple trials. The lower
mass cut star we kept in G180.0$-$01.7 has a 14\% false positive 
probability and an intercept time that is too old.  G260.4$-$03.4
has a large number of lower mass candidates, all of which have very
high false positive rates.  Many of the parallaxes imply distances 
that are likely behind the SNR, although several have intercept times
consistent with estimates for the age of the SNR.  G296.5$+$10.0 has
two candidates with high false positive probabilities but ages and
parallaxes more or less consistent with estimates for the SNR.  G130.7$+$03.1
and G263.9$-$03.3 have no candidates even for the $>1M\odot$ mass cut.

Our various false positive tests make it clear that you will find candidates
associated with any SNR if you search enough stars.  The candidates
in Table~\ref{tab:candbin} come from going to lower and lower mass limits
until there are candidates.  They almost all have high false positive
probabilities, so they are almost all likely to be false positives other
than HD~37424.  However, with the available data and uncertainties, 
we cannot definitively prove this for particular stars.  What we can
conclude is that in this sample of ten SNRs, there is only one unbound
companion with a mass $>5M_\odot$, as we do think it is reasonable to
exclude the two candidates associated with G069.0$+$02.7.  

The unbound star search for the three SNRs with binaries is really
a search for triple systems in which one star remains
bound to the compact object and the other star becomes unbound.  We
find no such candidates for G039.7$-$02.0 (SS~433), G205.5$+$00.5
(HESS~J0632$+$05 and MWC~148) or G284.3$-$01.8 (1FGL~J1018.6$-$5856
and 2MASS~J10185560$-$5856459).  For SS~433,
we carried out the search both for the Gaia EDR3 parallax and for the
distances estimated from kinematic models of SS~433 by \cite{Blundell2004}
and \cite{Marshall2013}.  Table~\ref{tab:searchbin} given the results
for the Gaia parallax distance.  The limits for SS~433 should only be regarded
as holding for $>5M_\odot$ due to its larger distance and extinction
compared to the other SNRs we consider.  We can clearly reject the
candidate HD~261393 found for G205.5$+$00.5 by \cite{Boubert2017}.  Its
Gaia EDR3 parallax of $\varpi = 0.923 \pm 0.025$~mas is inconsistent
with that of MWC~148 ($\varpi = 0.540\pm 0.023$~mas) at $\sim 130 \sigma$.  
The remaining \cite{Lux2021} candidates in this SNR also have parallaxes
incompatible with MWC~148.

\section{Disrupted Triples}

For the SNRs with single neutron stars, 
we can carry out the search for fully disrupted triples.  We now
keep candidates with $\chi^2_\varpi+\chi^2_\mu < 9.210$ and allow
pairings including all stars meeting the $>1M_\odot$ mass cut.
Because the two stars must have similar parallaxes and proper
motions that intersect within $10^5$~years and $R/R_{SNR}<0.2$,
the problem of false positives is greatly reduced over the search
for unbound binaries.  Nonetheless, it is still simply a question
of how low a mass limit can be used before you have significant 
probabilities of false positives, not whether you will have false
positives once you include enough stars.

The candidates are listed in Table~\ref{tab:candtrip}.  There are now
only four SNRs with candidates despite the relaxed mass limits.
Except for HD~37424, they all have high false positive probabilities,
and G069.0$+$02.7 again has the most.
Basically, since G069.0$+$02.7 has the largest number of stars which
are unbound binary candidates, it is also much more likely that two of these
candidates can have similar parallaxes and intercept times.  
Most of the candidates have rather high values for their
$\chi^2_\varpi+\chi^2_\mu$ fit statistic. 

Curiously, the best
candidate again involves HD~37424, where there is a 
much lower mass star ($M_{lim} >1M_\odot$) with a consistent
distance ($\chi^2_\varpi = 0.3$) and a good ($\chi^2_\mu=1.8$), 
simultaneous intercept with the NS $29$ thousand years ago close to the
center of the remnant ($R/R_{SNR}=0.1$).  It does, however, have
a significant false positive probability.  Its kinematics would
allow it to be in a wide binary with the neutron star progenitor and HD~37424.
In the local standard of rest (see below in \S5), its proper motion
corresponds to a velocity of $\sim 22$~km/s as compared to $\sim 95$~km/s
for HD~37424.  If we view these as circular velocities, and assume the
neutron star progenitor and HD~37424 had similar masses, this would make the 
semi-major axis of the outer binary $\sim 40$ times that of the inner
binary.  Of course, the actual ratio would depend on the orbital
phases at the time of explosion and the degree to which the system
lay in the plane of the sky.  There will, however, be a tendency for
slower moving stars coincidentally close to the explosion site to 
produce false positives because the actual direction of their motions
becomes relatively unimportant.  

Like the examples of unbound binary candidates, we cannot prove that
the unbound triple candidates in Table~\ref{tab:candtrip} are all false
positives, although it seems likely.  What we can say is that these
seven SNRs contain no examples of disrupted massive star triple systems.  Conservatively
none of these neutron stars was in a triple containing two $>5 M_\odot$ stars in
addition to the SN progenitor.

\begin{table*}
  \centering
  \caption{Luminous Stars Near HD~37424 and PSR~J0538$+$2817 }
  \begin{tabular}{lrccccl}
  \hline
  \multicolumn{1}{c}{Star} &
  \multicolumn{1}{c}{$\chi^2/N_{dof}$} &
  \multicolumn{1}{c}{$\log(T_*/\hbox{K})$} &
  \multicolumn{1}{c}{$\log(L_*/L_\odot)$)} & 
  \multicolumn{1}{c}{$M_*/M_\odot$} &
  \multicolumn{1}{c}{$\log t$} &
  \multicolumn{1}{c}{Comments} 
   \\
  \hline
    HD~39746 &$ 0.80 $ &$  4.281 \pm  0.040 $ &$  4.761 \pm  0.124 $ &$ 12.6 $-$ 18.2 $ &$ 6.99 $-$ 7.22 $ & B1II \\ 
   V1164~Tau &$ 1.43 $ &$  4.344 \pm  0.050 $ &$  4.524 \pm  0.151 $ &$  9.9 $-$ 16.5 $ &$ 7.00 $-$ 7.39 $ & A, Ell \\ 
    HD~38658 &$ 0.68 $ &$  4.179 \pm  0.033 $ &$  4.424 \pm  0.107 $ &$  9.9 $-$ 13.6 $ &$ 7.17 $-$ 7.39 $ & B3II \\ 
    HD~36665 &$ 1.02 $ &$  4.324 \pm  0.052 $ &$  4.379 \pm  0.163 $ &$  8.8 $-$ 14.9 $ &$ 7.05 $-$ 7.49 $ & B1, Be star \\ 
   HD~38017A &$ 0.84 $ &$  4.258 \pm  0.027 $ &$  4.289 \pm  0.066 $ &$  9.7 $-$ 11.6 $ &$ 7.28 $-$ 7.42 $ & B3V, visual binary \\ 
    HD~37424 &$ 0.89 $ &$  4.431 \pm  0.044 $ &$  4.219 \pm  0.128 $ &$ 10.1 $-$ 15.4 $ &$< 7.38$          & B9 \\ 
    HD~35347 &$ 1.61 $ &$  4.220 \pm  0.032 $ &$  3.837 \pm  0.081 $ &$  7.0 $-$  8.7 $ &$ 7.49 $-$ 7.69 $ & B1V, Be star \\ 
      ET~Tau &$ 1.22 $ &$  4.160 \pm  0.050 $ &$  3.819 \pm  0.160 $ &$  6.2 $-$  9.4 $ &$ 7.44 $-$ 7.81 $ & B8, EB \\ 
   HD~246821 &$ 0.84 $ &$  3.886 \pm  0.043 $ &$  3.715 \pm  0.142 $ &$  5.8 $-$  8.9 $ &$ 7.48 $-$ 7.90 $ & F0 \\ 
   HD~248666 &$ 0.82 $ &$  4.172 \pm  0.057 $ &$  3.708 \pm  0.167 $ &$  5.7 $-$  8.8 $ &$ 7.50 $-$ 7.89 $ & B \\ 
    HD~38749 &$ 0.75 $ &$  3.897 \pm  0.023 $ &$  3.599 \pm  0.072 $ &$  6.1 $-$  7.4 $ &$ 7.64 $-$ 7.85 $ & A5 \\ 
   HD~247176 &$ 0.29 $ &$  4.216 \pm  0.070 $ &$  3.536 \pm  0.204 $ &$  4.9 $-$  8.5 $ &$ 6.69 $-$ 8.04 $ & B2V \\ 
   HD~244610 &$ 1.46 $ &$  4.076 \pm  0.073 $ &$  3.427 \pm  0.197 $ &$  4.6 $-$  7.5 $ &$ 7.63 $-$ 8.12 $ & B1V, Be star \\ 
 TIC~3219130 &$ 0.91 $ &$  3.528 \pm  0.002 $ &$  3.337 \pm  0.014 $ &$  0.6 $-$  2.7 $ &$ 8.84 $-$ 9.99 $ & M5 \\ 
   HD~246370 &$ 0.17 $ &$  3.831 \pm  0.022 $ &$  3.314 \pm  0.068 $ &$  5.3 $-$  6.3 $ &$ 7.79 $-$ 7.96 $ & G5 \\ 
TIC~76026324 &$ 1.20 $ &$  3.587 \pm  0.005 $ &$  3.295 \pm  0.013 $ &$  1.6 $-$  5.6 $ &$ 7.91 $-$ 9.37 $ & M2/3 \\ 
   HD~245247 &$ 0.81 $ &$  3.816 \pm  0.012 $ &$  3.073 \pm  0.026 $ &$  4.7 $-$  5.3 $ &$ 7.97 $-$ 8.09 $ & A7 \\ 
    V399~Aur &$ 1.79 $ &$  3.529 \pm  0.002 $ &$  2.856 \pm  0.016 $ &$  0.7 $-$  1.3 $ &$ 9.66 $-$ 9.99 $ & M2/3, LPV \\ 
  \hline
  \end{tabular}
  \label{tab:stars}
\end{table*}

\section{HD~37424 and PSR~J0538$+$2817 }
\label{binary}

\begin{figure}
\centering
\includegraphics[width=0.50\textwidth]{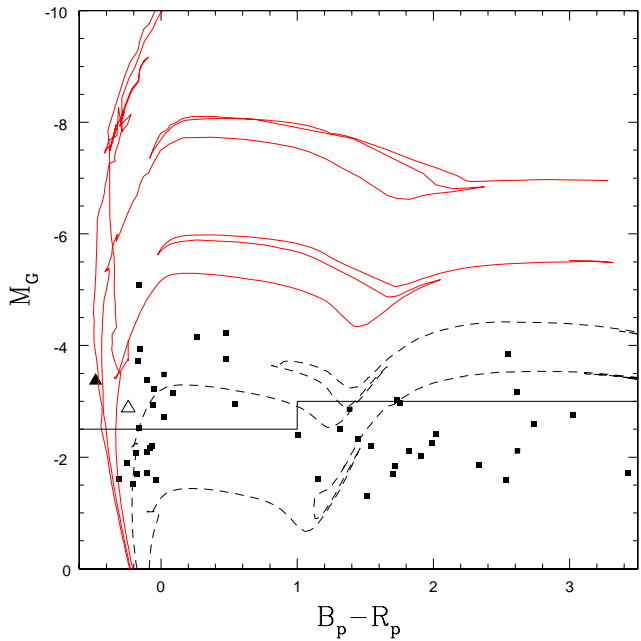}
\caption{
  Extinction-corrected CMD of stars with $G<10$~mag within $\pm 200$~pc
  along the line-of-sight and $100$~pc in the plane of the sky of 
  G180.0$-$01.7.  The large filled (open) triangle is HD~37424 based the
  {\tt mwdust} (SED fit) extinction estimate.  The red solid PARSEC
  isochrones are for $\log t = 6.6$, $7.0$ and $7.5$, while
  the black dashed isochrones are for $\log t = 8.0$ and $8.5$.  The
  maximum initial masses for stars remaining on the $\log t = 7.0$, $7.5$, 
  $8.0$ and $8.5$ isochrones are $19.3$, $9.1$, $5.3$ and $3.4M_\odot$,
  respectively.  The SED models for the 18 stars above the horizontal
  lines are presented in Table~\protect\ref{tab:stars}.
  }
\label{fig:g180cmd}
\end{figure}

Fig.~\ref{fig:g180cmd} shows a Gaia color-magnitude diagram (CMD) of
all stars with parallaxes of $0.581 < \varpi < 0.758$~mas, which is 
nominally $\pm 200$~pc around the
joint parallax ($\varpi = 0.658 \pm 0.028$~mas, or $1.52$~kpc)
of HD~37424 and the neutron star, angles $< 3.77^\circ$ from the
center of the SNR, which is $100$~pc at the nominal distance, and
with $G<10$~mag (1~mag fainter than HD~37424).  This yielded a sample
of $50$~stars.  If we compare these stars to the
PARSEC isochrones with $\log t = 6.6$, $7.0$, $7.5$, $8.0$ and $8.5$,
we see that they are all consistent with $\log t \gtorder 7.5$, and
the initial mass of the most massive stars remaining on this isochrone
is $9.1M_\odot$.  There are certainly no stars close to the $\log t=7.0$
isochrone where the initial mass of the most massive remaining star
is $19.3 M_\odot$.

At the absolute magnitude of HD~37424, the $\log t \ltorder 7$ isochrones
are nearly vertical, so any estimate of its age is largely determined by
its color.  \cite{Dincel2015} spectroscopically classified HD~37424 as a 
B0.5V star, and estimated a mass and luminosity of $M_* \simeq 13 M_\odot$ and
$L_* \simeq 14000 L_\odot$ based on the calibrations
of spectral types by \cite{Hohle2010}.  

We can construct a 
spectral energy distribution (SED) for HD~37424 spanning from 1570\AA\ 
to $4.6\mu$m using UV fluxes from \cite{Thompson1978}, optical fluxes from
NOMAD (\citealt{Zacharias2005}), near-IR fluxes from 2MASS (\citealt{Skrutskie2006}),
and mid-IR fluxes from AllWise (\citealt{Cutri2012}).  We used
a temperature prior of $T_* = 26000 \pm 3000$~K based on the spectral
type, and an extinction prior of $E(B-V) = 0.5 \pm 0.1$
based on the {\tt mwdust} extinction estimate. The photometric errors were
set to be the larger of the reported uncertainties and 10\% with the exception
of the 2740\AA\ UV flux. The error for this band was set to 30\% because it
markedly disagrees with any smooth SED.  The minimum photometric uncertainty
is to compensate for systematic uncertainties.  We fit the SED using Markov
Chain Monte Carlo methods and Solar metallicity \cite{Castelli2003} atmosphere models to find
that $T_* = 27000 \pm 2600$~K, $L_* = 16600 \pm 4500 L_\odot$, 
$R_* = 5.9 \pm 0.3 R_\odot$ and $E(B-V) = 0.34 \pm 0.04$.  The goodness of fit is
reasonable, with $\chi^2 = 9.7$ for $11$ degrees of freedom (12 photometric
measurements, 2 priors and 3 variables).  The SED slightly constrains the
temperature over the prior from the spectral type (B9). The SED requires less
extinction than the {\tt mwdust} estimate, which shifts the star redwards to
colors more typical of the other nearby massive stars (see Fig.~\ref{fig:g180cmd}). 

If we search for models with these luminosities and temperatures in the Solar
metallicity {\tt PARSEC} isochrones we can estimate the mass and age of the
system.  We selected models using a $\chi^2$ statistic to measure the distance
of the model from the estimated stellar parameters and simply tracked the 
maximum and minimum values within a given $\chi^2$ limit.  
In matching to the isochrones 
we used the larger of the estimated uncertainties and 5\% or 10\% for 
the temperature and luminosity, respectively.
For $\chi^2<1$,
the current mass and age ranges were $11.1 M_\odot < M_* < 13.9 M_\odot$ and
$\log t < 7.18$, while for $\chi^2 < 4$, the ranges were
$10.0 M_\odot < M_* < 15.3 M_\odot$ and $\log t < 7.38$.  There was no useful 
lower limit on the age.  The upper age
limits are certainly consistent with the star being the companion of
a supernova, as the maximum initial masses for the $\log t = 7.18$ and $7.38$
isochrones are $14.0 M_\odot$ and $10.6 M_\odot$, respectively.  
These parameters are summarized in Table~\ref{tab:stars}.

\begin{figure}
\centering
\includegraphics[width=0.50\textwidth]{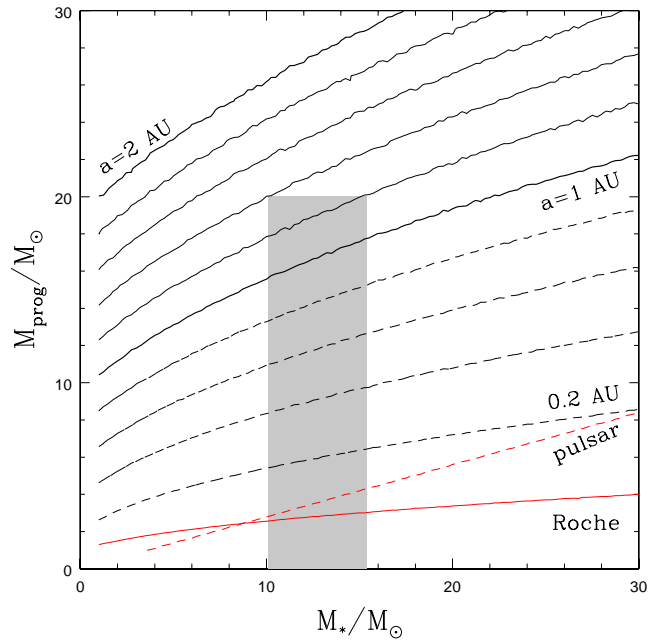}
\caption{
  Semimajor axes as a function of the mass of HD~37424 ($M_*$) and the
  mass of the SN progenitor at the time of explosion ($M_p$).  The
  shaded grey region is allowed by the fits to the SED of HD~37424 and
  the ages of nearby massive stars for the progenitor.  Below the 
  red Roche line, HD~37424 would fill its Roche lobe, and below the
  red dashed pulsar line the neutron star kick amplitude and direction
  become improbable.  The kinematic uncertainties allow a $35\%$ dispersion
  in the semimajor axis at fixed mass.  There is a narrow band of 
  acceptable pulsar solutions below the Roche line which has been
  excluded for clarity.  
  }
\label{fig:orbit}
\end{figure}

\begin{figure}
\centering
\includegraphics[width=0.50\textwidth]{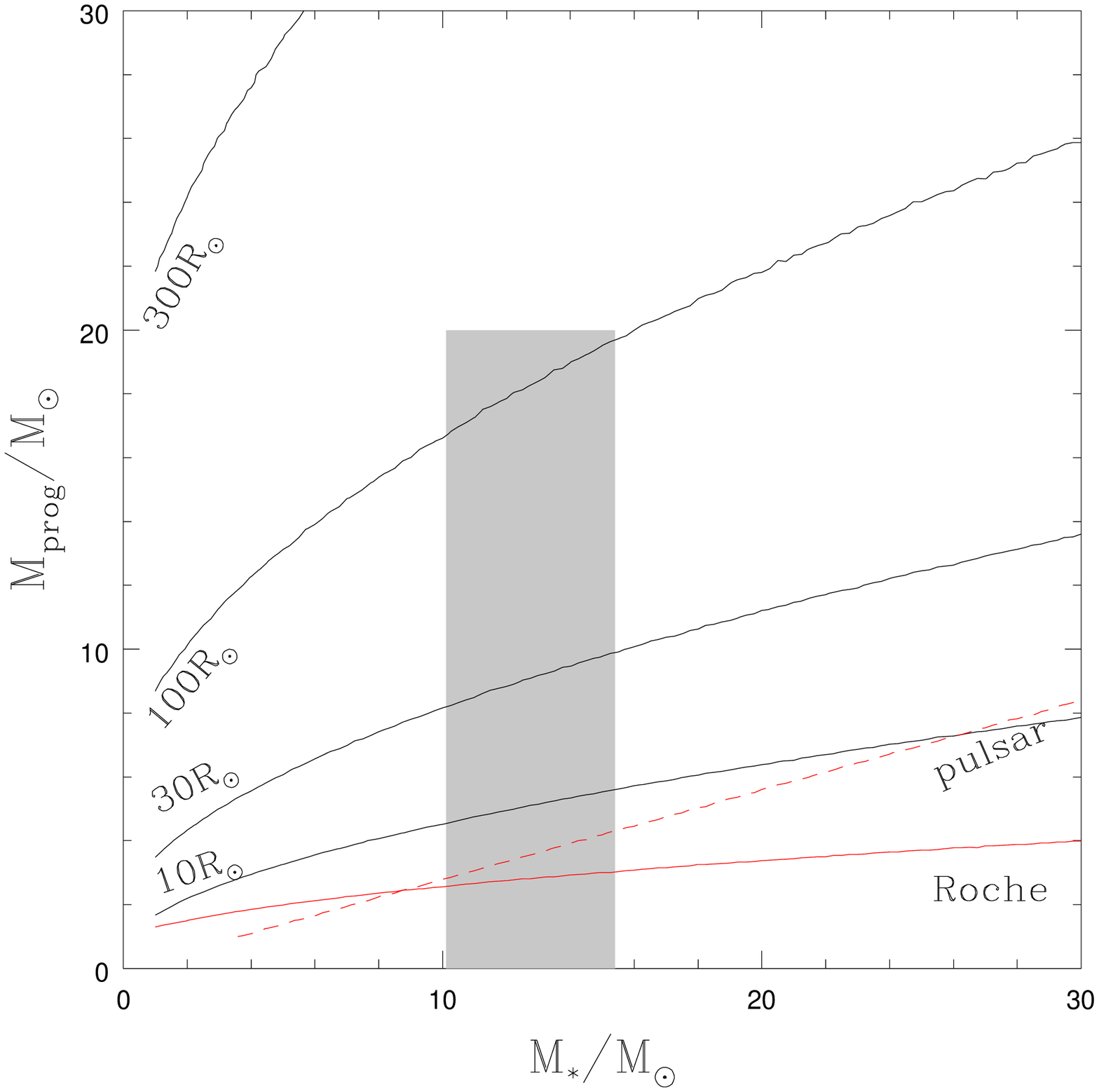}
\caption{
  The Roche limit on the SN progenitor (black contours) 
  over the same mass range as in Fig.~\ref{fig:orbit} and with
  the same limits from the Roche radius of HD~37424 and the
  pulsar kinematics. The Roche limit on the progenitor is
  much smaller than the radii of red supergiants.  The 
  most likely scenario was that the progenitor was Roche
  lobe filling, transferring mass which ``rejuvenated''
  HD~37424 to make it appear younger than other nearby
  massive stars and then leading to a Type~IIL or IIb
  supernova. While not drawn, a $3R_\odot$ Roche radius for
  the SN progenitor closely tracks the Roche radius limit   
  for HD~37424.
  }
\label{fig:roche}
\end{figure}

We also modeled the SEDs of the 18 stars with {\tt mwdust} extinction-corrected 
absolute magnitudes of $M_G<-2.5$ ($M_G < -3.0$) and
$B_P-R_P < 1$ ($>1$).  The limit on the blue main sequence is
faint enough to include all stars which might eventually undergo
core collapse, and the modest shift upwards in luminosity 
for redder colors was designed so that a few lower mass
red giants were included to provide a check that they 
were lower mass without having to model many of them. This
selection limit is shown in Fig.~\ref{fig:g180cmd}.  
For the red giants,
we generally used APASS (\citealt{Henden2016}) instead of NOMAD for the optical
magnitudes and the MARCS (\citealt{Gustafsson2008}) atmosphere models.
We had to broaden the photometric uncertainties of some
bands to 20-30\% to reach a reasonable $\chi^2/N_{dof}$.
All the hotter stars had UV fluxes from \cite{Thompson1978}, which provided 
enough temperature information to make it clear that the hottest
stars were B and not O stars, consistent with the reported spectral 
types.  This also allowed very robust individual extinction estimates.
Table~\ref{tab:stars} reports the resulting temperatures and luminosities
along with the $\chi^2<4$ range of PARSEC isochrone initial masses and ages
to which they could correspond.  
The luminosities, temperatures, masses and ages of these stars,
along with their spectral types and any interesting properties
in the {\tt SIMBAD} (\citealt{Wenger2000}) database, are
also given in Table~\ref{tab:stars} and sorted by luminosity.

Most of the stars shift very little in the CMD with the directly
estimated extinctions.  The exception is TIC~3219130 where the
{\tt mwdust} extinction estimate of $E(B-V) = 0.98$ is much too
high -- the SED implies $E(B-V)=0.31$ which makes the star a
rather low mass red giant.  The other two red giants (V339~Aur
and TIC~76026324) are also relatively low mass, consistent with
their locations on the CMD. The temperature estimates from the SEDs
are generally consistent with the spectroscopic classifications
with the exception of V1164~Tau. 

 {\tt SIMBAD} flags three stars
as variables, V1164~Tau, ET~Tau and V399~Aur.  The classification
of V1164~Tau as an ellipsoidal variable (ell) is from an automated 
classifier with no detailed analysis. As a luminous red giant,
it is not surprising that V399~Aur is a long period variable.
ET~Tau is a fully analyzed eclipsing binary (EB), where \cite{Williamon2016}
report parameters of $M_1 \simeq 14.3$, $T_1 \simeq 30300$~K, $L_1 =30500L_\odot $,
and $M_2=6.3 M_\odot$, $T_2=15000$~K and $L_2=6400L_\odot$, respectively.
This mass for the primary is much higher than found here ($5.8$-$8.9M_\odot$),
but the SED also clearly rules out the high temperature and luminosity of
the \cite{Williamon2016} model.
In particular, the SED clearly breaks going into the UV rather than continuing the 
Rayleigh-Jeans rise that would be expected for $30000$~K star.

Aside from HD~37424, all of the stars for which the age is constrained
(i.e., excluding stars like HD~247176) favor ages $\log t \gtorder 7.0$, and
most are closer to $\log t =7.5$.  There are clearly a number of stars
which have masses $>10M_\odot$, so it seems likely that the progenitor
of PSR~J0538$+$2817 had an initial mass between $\simeq 13M_\odot$, since there are 
many remaining stars in this mass/age range, and $19M_\odot$ since all
the stars appear to be older than $\log t=7$.  \cite{Dincel2015}
argued for a mass range of $13 M_\odot$, their estimate of the mass
of HD~37424 based on the spectral type, up to $25 M_\odot$ based
on the absence of nearby O~stars.  Note that we are not using
HD~37424 in setting this limit because we will next argue that  
the system must have been a mass transfer binary.

We would like to determine the properties of the pre-SN binary to constrain
both the size of the orbit and the mass of the progenitor.
Unfortunately, the geometry of the positions and proper motions can at best
set a minimum projected size for the binary.  The first problem
is simply that parallaxes provide no useful information for constraining the line
of sight separation at the time of the SNe.  Then, in projection, it is only
possible to rule out small separations, but not to prove them.  A solution
with an explosion at a projected separation of zero is allowed if there
is a reasonable probability that the cross product of the present day
separation and proper motion differences can be zero.  This is true for
HD~37424, although it need not be so.  However, the raw probability 
of a given minimum projected separation $R_\perp$ is just
proportional to $R_\perp$ because given a random choice for $R_\perp$ 
a value close to zero is unlikely. A Bayesian corrects for this by using
logarithmic prior of $1/R_\perp$, leading to a probability 
distribution which is independent of $R_\perp$ when it
is small. For sufficiently large
$R_\perp$ ($\sim 30000$~AU for HD~37424), the distribution cuts off
because large projected separations are unlikely. It is, after all,
how the star was selected.   However, while we select based on the
minimum projected separation, the explosion need not happen at this
closest approach distance, so a larger separation cannot be ruled out.   
In sum, the positions and proper motions would allow the progenitor system
to be anything from a contact binary to a very, very wide binary.

It is really the velocity of the star which constrains the orbit, which
is why \cite{Dincel2015} focus on the projected velocity of the star. 
They estimate a rest frame proper motion of ($10.0\pm 0.8$, $-5.9 \pm 0.6$)~mas/year after 
correcting for Galactic rotation and Solar motion using a model.  For the Gaia EDR3 parallax distance 
of $1.5$~kpc, this corresponds to $87\pm 9 $~km/s.  They also measured an average
heliocentric radial velocity of $-8.9\pm 3.0$~km/s.  This implies an
orbital separation smaller than the radius of a massive red supergiant,
so \cite{Dincel2015} suggest that the progenitor was a naked helium
star with a mass as low as $2 M_\odot$. 

With Gaia, we can use the proper motions and radial velocities
of nearby stars to define the local standard of rest.
We selected stars in the same parallax range used for the luminous stars, a 
larger angular separation ($10^\circ$) and apparent magnitudes within $1.5$~mag of HD~37424
since kinematics depend on mass/luminosity.  
This yielded 523 stars with proper motions and
345 with radial velocities.  HD~37424 does not have a Gaia radial velocity.
These stars had a median
proper motion of ($0.78$, $-3.3$)~mas/year with a dispersion (estimated
from the central 68\% of the sorted proper motions)
of ($1.9$, $2.4$)~mas/year, and a median radial velocity of 
$0.1$~km/s with a dispersion of $29.8$~km/s.  Note that the
intrinsic dispersions are more important than the typical
measurement uncertainties.
Subtracting this from the proper motion of HD~37424 implies an SN
local rest frame proper motion of ($11.5$, $-6.1$)~mas/year with uncertainties
set by the local proper motion dispersion, or $94 \pm 15$~km/s.  Similarly,
subtracting the mean local RV from the \cite{Dincel2015} velocity for
HD~37424 gives a local rest frame radial velocity of $-9.0 \pm 30.0$~km/s.
The larger dispersion in the radial velocity is puzzling.

The star has a small radial velocity compared to the velocities in the plane of
the sky, and there are two arguments that the same is true of the present neutron
star motions.  First, \cite{Ng2007} argue that the pulsar wind nebula 
is seen almost edge on and aligned nearly perpendicular to the proper
motion.  Second, \cite{Yao2021} use
scintillation of the pulsar signals off the near side of the SNR 
to estimate its distance from the neutron star and argue that the estimated
distance puts the neutron star close to the center of the SNR and so 
implies that it has a low line-of-sight velocity.
For the neutron star this could be entirely due to the kick, but it
seems contrived to have the kick nearly cancel the motions along the 
line of sight.  So for the
present discussion, we assume that the pre-supernova binary effectively lay in the 
plane of the sky and that the orbit was circular.  

Under these assumptions,
the velocity of the star in the plane of the sky is
\begin{equation}
  \vec{v}_* = v_o { M_p \over M_T} \left( \begin{matrix} \cos\theta \\ \sin\theta\\ \end{matrix} \right)
\end{equation}
and the velocity of the neutron star is  
\begin{equation}
  \vec{v}_{NS} = - v_o { M_* \over M_T} \left( \begin{matrix} \cos\theta \\ \sin\theta\\ \end{matrix} \right)
     + v_k \left( \begin{matrix} \cos\psi \\ \sin\psi\\ \end{matrix} \right)
\end{equation}
where $M_*$ is the mass of HD~37424, $M_p$ is the mass of the SN progenitor at death,  $M_T = M_* + M_p$ is the
total mass, $v_0=(GM_T/a)^{1/2}$ where $a$ is the semimajor axis, $v_k$ is the neutron star kick velocity 
and $\theta$ and $\psi$ are the direction of the stellar velocity and the kick, respectively.
As a function of the two masses, the proper motions and the estimated local standard of rest
combine to give estimates of $a$, $v_k$, $\theta$ and $\psi$.  

We have several additional constraints on these parameters.  First, from the SED models we
have an estimate of $M_*$.  Second, the pulsar spin axis, generally believed to be perpendicular to the wind nebula, 
is nearly perpendicular to the proper motion of the pulsar, at a PA of $\theta_{spin}=-11^\circ$ or $12^\circ$ 
off the neutron star velocity vector (\citealt{Ng2007}).  This alignment is common and \cite{Ng2007b}
use the distribution of mis-alignments to try to constrain the origin of neutron star kicks.  If we model
the mis-alignments reported by \cite{Ng2007b} with a Gaussian distribution corrected for the
individual uncertainties, we find a best fit dispersion of $\sigma_\psi = 17^\circ$.  So
we can constrain the angle $\psi$ as a position angle to be within $\sigma_\psi$ of 
$\theta_{spin}$.  Third, there are constraints on the amplitudes of neutron star kick velocities,
where we use the estimate that the distribution of 1D coordinate velocities 
can be modeled as a Gaussian with a dispersion of $\sigma_k = 265$~km/s (\citealt{Hobbs2005}). 
Finally, we also have an estimate of the radius of HD~37424 from the SED models and can
eliminate mass ranges where the Roche radius is smaller than the stellar radius,
where we use the approximation for the mean Roche radius of \cite{Eggleton1983}.

Fig.~\ref{fig:orbit} summarizes the results in the plane of the mass of HD~37424, $M_*$,
and the mass of the progenitor at death, $M_p$. A broader than plausible mass range is shown for both,
with a gray shaded region giving the SED-based mass estimates for HD~37424 and a generous upper
limit of $M_p < 20 M_\odot$ based on the properties of the surrounding stars.  Contours
show the semi-major axis of the orbit implied by the masses and kinematics.  For a given mass,
the uncertainties in the kinematics produce a dispersion of $\simeq 35\%$ around this central
value across the plane of masses.  For progenitor masses below the line labeled ``Roche'',
HD~37424 would be overflowing its Roche lobe given its radius of $(5.9 \pm 0.3)R_\odot$.
For progenitor masses below the line labeled ``pulsar,'' the neutron star properties
(kick and alignment) are unlikely.  Over the rest of the plane, both the implied
kick velocity and its alignment relative to the spin are consistent with expectations.
There is a narrow band below the Roche curve where the pulsar kick velocity and direction are again consistent
with expectations that we have suppressed for clarity.

In terms of the orbital kinematics, production of a neutron star in a supernova  and 
the properties of the neutron star kick, there is a broad progenitor mass range 
leading to reasonable solutions.  The problem, already noted by \cite{Dincel2015},
is that a $10$-$20M_\odot$ red supergiant should be significantly larger than these
semimajor axes.  \cite{Groh2013} find $547R_\odot=2.5$~AU ($931R_\odot=4.3$~AU) for an initial mass of 
$10M_\odot$ ($20 M_\odot$) and \cite{Sukhbold2016} have $414R_\odot=1.9$~AU ($1065R_\odot=5.0$~AU). This
problem cannot be solved by inclining the orbit out of the plane of the sky. Fig.~\ref{fig:roche}
illustrates this problem with contours for the Roche limit of the progenitor.  \cite{Dincel2015}
argue that the progenitor may have been a naked helium star at death, but this pushes
into the very low pre-SN progenitor mass range where there are problems with the pulsar
kinematics and the Roche limit of HD~37424.  

The simplest solution is that the system was likely an interacting binary at death, with
a Roche lobe-filling progenitor transferring mass to HD~37424.  The mass transfer would
help to explain why HD~37424 appears anomalously young and hot compared to the other 
stars in the field.  The Roche limit is sufficiently small that the progenitor would
have lost much of its hydrogen leading to a Type~IIL
or IIb supernova.  In combination, these two classes represent $\sim16\%$ of ccSNe
(\citealt{Smith2011}, \citealt{Eldridge2013}), so it is 
not a particularly improbable scenario for a randomly selected case.

\begin{table*}
  \centering
  \caption{Constraints on Binaries}
  \begin{tabular}{lcccc}
  \hline
  \multicolumn{1}{c}{Case} &
  \multicolumn{2}{c}{Non-Interacting Incomplete}  &
  \multicolumn{2}{c}{Non-Interacting Complete  }  \\
  &
  \multicolumn{1}{c}{Median} &
  \multicolumn{1}{c}{90\% Confidence} &
  \multicolumn{1}{c}{Median} &
  \multicolumn{1}{c}{90\% Confidence} \\
  \hline
  Not Binary at Death     &72.0\% &52.2\%-86.4\% &76.0\% &56.0\%-89.2\% \\
  Bound Binary            &13.9\% &5.4\%-27.2\%  &9.5\%  &3.6\%-19.7\%  \\
  Interacting Binary      &10.2\% &3.2\%-22.3\%  &5.8\%  &1.8\%-13.1\%  \\
  Non-Interacting Binary  &       &$<$8.5\%      &       &$<$8.9\%      \\
  Unbound Binary          &12.5\% &2.8\%-31.3\%  &13.2\% &3.0\%-32.8\%  \\
  \hline
  \end{tabular}
  \label{tab:binary}
\end{table*}

\section{Discussion}
\label{sec:conclude}

We searched for unbound stellar companions to 7 non-binary neutron stars,
two binary neutron stars and one binary black hole system that are both
associated with SNRs and have proper motions.  We identify only one convincing
candidate, HD~37424 in G180.0$-$01.7, which was originally identified
by \cite{Dincel2015} and also confirmed using different methods by 
\cite{Boubert2017}. All the remaining candidates are lower mass
and have high probabilities of being false positives.  
The mass limit to which companions can be ruled out
depends on the SNR, but a conservative limit is that there are no 
unbound binary or triple systems other than HD~37424 with masses 
$\gtorder 5 M_\odot$. 

We estimate that HD~37424 has a mass of 
$10$-$15M_\odot$ and that it was probably in a mass transfer binary
with a semi-major axis of $\sim 1$~AU that would partly strip the 
neutron star progenitor and lead to a Type~IIL or IIb SN.  The progenitor
was likely a $13$-$19M_\odot$ star at birth based on the properties
of the stars within $100$~pc of the explosion.  These conclusions
agree with \cite{Dincel2015} except for the nature of the progenitor
at death, where \cite{Dincel2015} suggest it was a fully stripped
helium star.
    
By combining this work, \cite{Kochanek2018} and \cite{Kochanek2019}, we
can constrain all of the possibilities for the binary properties and
outcomes of explosions producing neutron stars.  In \cite{Kochanek2018}
we showed that the Crab, Cas~A and SN~1987A were not binaries
at the time of explosion.  Because these SNRs are so young, there
is no need for proper motion information.  We are assuming that 
SN~1987A produced a neutron star (see, e.g., \citealt{Cigan2019}, 
\citealt{Greco2021}).  In \cite{Kochanek2019} we searched for 
surviving binaries starting from an initial sample of 49 SNRS.
Of these, 23 contained compact objects where it was practical
to carry out a search, and this included three previously identified
HMXBs.  If we drop SS~433 as a black hole system, then there
were two interacting binaries and no non-interacting binaries in
a sample of 22 explosions producing neutron stars. 

We can combine all of these results to jointly estimate the fraction
$f_n=1-f_u-f_b$ of SN producing neutron stars that are not binaries at the time of the SN, the
fraction $f_u$ leading to binaries that are unbound after the explosion, the
faction $f_b=f_i+f_p$ leading to binaries that are bound, where fraction
$f_i$ produce interacting binaries and $f_p$ produce non-interacting
(passive) binaries.  The multinomial probability distribution is
\begin{equation}
    f_n^9 (1-f_b)^{11} f_u f_i^2 
\end{equation} 
since we have 9 SNR containing neither bound nor unbound binaries, 
11 which contain no bound binary but could be unbound binaries,
1 unbound binary and two interacting binaries.  Again, we 
conservatively mean stellar companions $\gtorder 5M_\odot$, 
although the mass limits for individual systems can be significantly
tighter.  This case, ``Non-Interacting Incomplete'' in Table~\ref{tab:binary},
assumes that the presence of the interacting companion does not
bias whether the interacting binary was included in the sample of
23 SNRs where \cite{Kochanek2019} searched for bound companions.
Using uniform priors for the fractions, the probability can be
marginalized to obtain the probability distribution for each of
these fractions, with the median values and 90\% confidence ranges
given in Table~\ref{tab:binary}.  

If, on the other hand, interacting binaries are so easy to detect
due to their X-ray, $\gamma$-ray or radio emission that the 23 SNRs
analyzed by \cite{Kochanek2019} included all interacting binaries
in the parent sample of 49 SNRs. If so, we have overestimated the
probability of producing an interacting binary.  If we assume
all interacting binaries were included (case ``Non-Interacting
Complete'') then there were 2 interacting neutron star binaries
in a sample of 43 SNRs after making a rough correction for
contamination by Type~Ia SN (see \citealt{Kochanek2019}).
In this case, the multinomial probability distribution becomes  
\begin{equation}
    f_n^9 (1-f_b)^{11} (1-f_i)^{20} f_u f_i^2 
\end{equation} 
because there are an additional 20 SNRs known not to contain
interacting binaries.  The results for this case are also given
in Table~\ref{tab:binary} and the true answer should be bounded
by these two limits.

We will not attempt a similarly complete sub-division 
into cases for the question of triple systems at death 
in part because we lack information on whether the 
interacting binaries and HD~37424 have distant bound
companions.
However, of the 10 systems where the compact object
is not in a binary
(the seven here plus the Crab, Cas~A and SN~1987A),
none are fully unbound triples. So of SN that do not leave
bound binaries, less than 18.9\% can be fully unbound
massive star triples.  Of the 3 systems which are
interacting binaries, none is a partially unbound
triple.  Thus, of SN leaving interacting binaries,
less than $<43.7\%$ can have an unbound triple 
companion.  Note that these are fractions of systems which are
not binaries or binaries after the explosion, while the 
statistics for binaries in Table~\ref{tab:binary} are 
all as fractions of SN.  

While there will be a point where systematic problems due to any
correlation between the detectable life times of SNRs and any of the
binary properties we consider will become important, 
statistical uncertainties are likely dominant at present (see the discussion
in \citealt{Kochanek2019}).  It certainly seems possible to
significantly improve the statistics simply by identifying more
compact objects in SNRs with reasonable distances and extinctions
and then measuring their proper motions.  In \cite{Kochanek2019}
we found that the problem for many of the SNRs which could be
studied was that there were multiple, faint, X-ray sources
superposed on the remnant.  Detecting a proper motion would
clearly separate the neutron star from the contaminating 
background sources.
Where the neutron star is already detected, a sufficient amount
of time has likely passed since the original observations that 
proper motions can be measured or at least tightly restricted.
Simply knowing that the neutron star has (say) moved less than
0\farcs1 in 20 years restricts the allowed position $10^4$ years
ago to under an arcminute.  

The biggest cause of false positives in searching for unbound 
stars is the lack of good distances to the SNRs.  In theory
this can be addressed by searching for absorption features 
associated with the SNR in the spectra of superposed stars. This
approach appears to have worked quite well for the Vela SNR,
where the estimated distance of $250 \pm 30$~pc by \cite{Cha1999}
based on the appearance of high velocity components in the
Na~D and Ca~II absorption lines of more distant stars
agrees almost exactly with the \cite{Dodson2003} pulsar
parallax distance ($\varpi^{-1} = 286 \pm 16$~pc).  The
absorption lines of the background stars to Vela are also
unusual because many of them are significantly time variable
(see \citealt{Rao2020} most recently).  With
Gaia distances and fiber-fed echelle spectrographs this
would appear to be a straight forward approach to measuring
the distances to many SNRs.  

Improving the constraints on fully unbound triples is much
easier than doing so for unbound binaries because accurate
distances to the SNR and pulsar proper motions are less 
needed for the control of false positives. Particularly
if restricted to relatively massive and rare stars (say $>5M_\odot$),
the requirement that two stars with Gaia proper motions and
parallaxes have consistent distances and paths that intersect
near the center of the SNR at some point in the last $10^5$
years will greatly suppress or eliminate false positives.
The same is also true for a partly unbound binary where
the compact object is in a binary because Gaia supplies
the parallax and proper motion of the binary companion to
the compact object.

HD~37424 is a very convincing case of an unbound binary companion,
but what can be done both to strengthen the evidence and use it to learn
something about binary evolution, supernovae and supernovae in
binaries? \cite{Dincel2015} looked for high velocity absorption
features in their echelle spectra of HD~37424 and found nothing
but note that the signal-to-noise ratio of the spectrum was 
relatively low.  Finding such features, would help to confirm the hypothesis
and it would be trivial to obtain far lower noise spectra from
the ground.  It might be better, however, to look in the UV with
HST where there are many more lines (see, e.g., the study of the
SN~1006 SNR by \citealt{Winkler2005}).
An unusually fast rotation rate would help to confirm the interacting
binary scenario, but a low estimate of $v \sin i$ will be difficult
to interpret given the evidence that the binary likely lay in the
plane of the sky.  The supernova shock wave affects the structure of 
the outer layers and the luminosity of the star, but any time evolution
from the relaxation back to steady state is likely impossible 
to detect with the passage of $~30,000$~years (see, e.g., \citealt{Ogata2021}). 
The best route may be to look for abundance anomalies, as several
HMXBs are known to have anomalous $\alpha$-element abundances presumably
due to capturing some of the ejecta from the SN (e.g., \citealt{Israelian1999},
\citealt{Orosz2001}).

\section*{Acknowledgments}

CSK thanks K. Belczynski for help with BSP models when this project
was headed in a somewhat different direction, and R. de Stefano, Jennifer
Johnson, M. Moe and M. Pinsonneault for various discussions.
CSK is supported by NSF grants AST-1814440 and AST-1907570.  
This work has made use of data from the European Space Agency (ESA) mission
{\it Gaia} (https://www.cosmos.esa.int/gaia), processed by the {\it Gaia}
Data Processing and Analysis Consortium (DPAC,
https://www.cosmos.esa.int/web/gaia/dpac/consortium). Funding for the DPAC
has been provided by national institutions, in particular the institutions
participating in the {\it Gaia} Multilateral Agreement.
 This research has made use of the SIMBAD database,
operated at CDS, Strasbourg, France

\appendix

\section{Sample}

We examined 10 SNRs where there is a proper motion measurement either
for the neutron star (7 systems) or its binary companion (3 systems) that also
have distances and extinctions small enough to search for stellar
companions using Gaia EDR3 parallaxes and proper motions.  For 
statistical analyses we also include the Crab (G184.6$-$05.8), 
Cas~A (G117$-$02.1) and SN~1987A. Because these SNRs are very
young, the plausible regions in which any former companion could lie
are so small that \cite{Kochanek2018} (also see, \citealt{Kerzendorf2019},
\citealt{Fraser2019}) could demonstrate that there were no bound
or unbound former companions down to mass ratios $\ltorder 0.1$
even in the absence of proper motion data.

\subsection{G039.7$-$02.0} 

  Contains the interacting compact object binary SS~433
  (for a review, see \citealt{Margon1984}).  We us the position and size of
  the SNR from \cite{Green2019}.  SS~433 is more distant ($\varpi^{-1} = 8.5$~kpc)
  and highly extincted ({\tt mwdust} $E(B-V)\sim 1$ to $2$~mag) than the other SNRs
  we consider here, so we selected stars
  to $G<20$~mag instead of $G<18$~mag.  Even so, this means we are really only
  sensitive to stars in the $M>5M_\odot$ bin.  To limit the number of stars
  given the faint magnitude limit,
  we only considered stars with $0 < \varpi \leq 0.25$~mas,
  corresponding to the $\sim 5\sigma$ range of the Gaia EDR3 parallax.
  The distance implied by the Gaia EDR3 parallax ($6.1$ to $13.9$~kpc at
  $2\sigma$) significantly disagrees with distance estimates from
  kinematic models of the SS~433 jets ($5.5\pm0.2$, \citealt{Blundell2004},
  $4.5\pm0.2$~kpc, \citealt{Marshall2013}) but the parallax $RUWE=1.24$
  is not anomalously high.  As a precaution, we also searched for unbound
  stars using a parallax of $0.208 \pm 0.042$~mas, which encompasses the
  kinematic distance estimates and again found no candidates.
  The age of the remnant is estimated to be anywhere
  from $10^4$ to $10^5$~years (see, e.g., \citealt{Panferov2017}).

\subsection{G069.0$+$02.7 (CTB~80)}

Contains the radio pulsar PSR~B1951+32 (\citealt{Kulkarni1988}). which
has a proper motion of ($-28.8\pm0.9$,$-14.7\pm0.9$)~mas/year (\citealt{Zeiger2008}).
There is no parallax for the pulsar.   \cite{Koo1993}
roughly estimate the distance to CTB~80 to be $d\simeq 2$~kpc, while
\cite{Leahy2012} estimate a distance of $1.5_{-0.4}^{+0.6}$~kpc. \cite{Kulkarni1988}
estimate $1.4$~kpc from the dispersion measure to the pulsar.  We adopt
a loose limit on the ``parallax'' of the SNR of $\varpi = 0.75 \pm 0.25$~mas
which encompasses $1$ to $2$~kpc at $1\sigma$.
We adopt 
19:54:50 33:00:30 for the center of the SNR from \cite{Zeiger2008},
and a diameter of 80~arcmin from \cite{Green2014}. The pulsar has a spin-down
age of $\tau_c = P/2\dot{P} = 107$ thousand years (\citealt{Fruchter1988}).
\cite{Koo1990} estimate an age from the size and expansion velocity of the
SNR of $77$~thousand years for a distance of $2$~kpc.
For a limiting magnitude of $G<18$~mag, the sample of $\sim M_\odot$ stars
will be incomplete, but the false positive rate is also too high to
search for such low mass companions.

\subsection{G109.1$-$01.0}

This SNR contains the  anomalous X-ray pulsar 1E~2259$+$586 (\citealt{Gregory1980}) which has
a proper motion of ($-6.4\pm 0.6$, $-2.3\pm 0.6$)~mas/year (\citealt{Tendulkar2013}).
\cite{Verbiest2012} estimate a
distance of $4.1\pm0.7$~kpc 
\cite{Kothes2012} 
and \cite{Sanchez2018} argue for a lower distance of $3.2 \pm 0.2$~kpc.
We adopt a ``parallax'' of $\varpi = 0.28 \pm 0.05$ which spans $3.2$
to $4.3$~kpc at $1\sigma$.
\cite{Kothes2006}
derive a center of 23:01:39 $+$58:53:00 with a diameter of 33~arcmin. 
The spin down age is $230$~thousand years, but estimates based on
the remnant are generally 10 to 20 thousand years (see, \citealt{Sanchez2018}).
$N(H)=(0.93\pm0.04) \times 10^{22}$~cm$^2$ (\citealt{Patel2001}),
corresponding to $E(B-V)=1.60\pm0.07$.  For a limiting magnitude of $G<18$~mag
and a distance of $4$~kpc we can only search for $>4M_\odot$ companions.

\subsection{G130.7$+$03.1 (3C58, SN~1181)}

  This SNR contains the X-ray pulsar PSR~J0205$+$6449 (\citealt{Murray2002}),
  which has a proper motion of ($-1.40\pm0.16$, $0.54\pm 0.58$)~mas/year
  \cite{Bietenholz2013}. 
  \cite{Roberts1993} estimate a kinematic (HI) distance
  of $3.2$~kpc, while \cite{Camilo2002} obtain $4.5_{-1.2}^{+1.6}$~kpc based on the
  dispersion measure to the pulsar, and \cite{Kothes2012} argues for $2$~kpc.
  The age is also uncertain. The spin down age is $5.4$~thousand
  years (\citealt{Livingstone2009}) while it would only be 0.84 thousand
  years old if associated with SN~1181 (\citealt{Clark1977}). Studies of
  the remnant (e.g., \citealt{Bietenholz2006}) generally argue that the
  SNR must be several thousand years old.
  We adopt a ``parallax'' of $0.36 \pm 0.14$~mas which spans $2.0$ to $4.7$~kpc
  at $1\sigma$.
  \cite{Gotthelf2007} propose 02:05:33.97 64:49:50 as the explosion 
  center\footnote{The text of \cite{Gotthelf2007} gives the declination
  of the center as 63:49:50, but this appears to be a misprint given
  the figures and the discussion.} and
  the asymmetric remnant has an average diameter of approximately 8.1~arcmin (\citealt{Reynolds1988}).
 \cite{Slane2004} find $N(H)=(4.5\pm0.1)\times 10^{21}$~cm$^2$,
  corresponding to $E(B-V) \simeq 0.77 \pm 0.02$. For a limiting
  magnitude of $G<18$~mag and a distance of $5$~kpc we can only
  search for $>3M_\odot$ companions.

\subsection{G180.0$-$01.7}

This SNR contains the radio pulsar PSR~J0538$+$2817 (\citealt{Anderson1996}).
 \cite{Ng2007} measured VLBI parallax and proper motions of $\varpi = 0.68\pm0.15$~mas
and ($-23.53\pm0.16$, $52.59\pm0.13$)~mas/year, and \cite{Chatterjee2009}
measured $\varpi=0.72_{-0.09}^{+0.12}$ and ($-23.57\pm0.10$,$52.87_{-0.10}^{+0.09}$)~mas/year.
The \cite{Chatterjee2009} analysis appears to use the \cite{Ng2007} data but
considers more sources of systematic error, so we use their values and
the larger values for the asymmetric errors.
\cite{Kramer2003} estimate that the center of the SNR is at
05:40:01 $+$27:48:09, and \cite{Dincel2015} estimate that the
diameter is $200$~arcmin.  Based on the disrupted binary, \cite{Dincel2015}
estimate an age of $(30\pm 4)$~thousand years.  The pulsar has 
a spin down age of $620$~thousand years (\citealt{Lewandowski2004}).
Based on a Sedov model of the SNR, \cite{Sofue1980} estimated the
age of the SNR as $200$~thousand years for a distance of $1.6$~kpc.
\cite{Ng2007} also find $N(H) \simeq (2.7\pm0.3)\times 10^{21}$~cm$^2$,
implying $E(B-V) \simeq 0.5 \pm0.1$.  Even at a higher distance of
$1$~kpc we can search for companions down to $>1M_\odot$.

\cite{Dincel2015} and later \cite{Boubert2017} identify HD~37424 as a 
candidate disrupted binary companion. HD~37424 has a Gaia EDR3 
parallax and proper motion of $\varpi=0.654 \pm 0.028$~mas 
and ($12.229\pm0.036$, $-9.407\pm 0.017$)~mas/year.  The parallaxes
are mutually consistent to $0.5\sigma$.
\cite{Dincel2015} were concerned that the stars HD~36665 and HD~37318 show 
absorption features due to the SNR but had distance estimates that placed them
in the foreground. In Gaia EDR3, these stars have parallaxes of
$\varpi = 0.702 \pm 0.031$~mas and $\varpi = 0.566 \pm 0.030$~mas, so
statistically, HD~36665 could be more distant, and HD~37318 clearly is
more distant than the SNR.

\subsection{205.5$+$00.5 (Monoceros Loop)}

  The Monoceros loop is associated with the
  $\gamma$-ray source HESS~J0632$+$057.  The SNR was identified
  prior to the discovery of the $\gamma$-ray source.
  \cite{Hinton2009} also
  identified it as an X-ray source and suggested that it was in a
  binary with the massive star MWC~148.
  The high energy emissions are believed to be due to
  interactions between a pulsar and the stellar wind
  (see the review by \citealt{Dubus2013}).
  We use the center and size of the SNR from \cite{Green2019}.
  \cite{Boubert2017} identify HD~261393 as a candidate disrupted binary
  companion in this SNR but its parallax is inconsistent with that
  of MWC~148 (see text).  \cite{Welsh2001} estimate an age of 30 to
  150 thousand years.

\subsection{G260.4$-$03.4 (Puppis A)}
 
  Puppis~A contains the X-ray pulsar PSR~J0821$-$4300 (\citealt{Gotthelf2009})
  where \cite{Mayer2021} measure a proper motion of 
   ($-74.2_{-7.7}^{+7.4}$, $-30.3_{-6.2}^{+6.2}$)~mas/year.  We
  round the asymmetric uncertainties upwards
  \cite{Reynoso2017} estimate a kinematic distance to the SNR of
  $1.3 \pm 0.3$~kpc, although earlier estimates in \cite{Reynoso2003}
  were higher, at $2.2$~kpc.   We adopt a ``parallax'' of
  $\varpi =0.75 \pm 025$~mas which spans $1$ to $2$~kpc
  at $1\sigma$.  
  \cite{Winkler1988} found an explosion center of 08:22:27.5
  $-$42:57:29 based on the proper motions of optical filaments
  in the SNR and \cite{Green2014} gives a diameter of 60~arcmin.
  The motions of the optical filaments also imply an age
  of $3700 \pm 300$~years (ignoring deceleration).
\cite{Hui2006} find $N(H)=(3.7\pm0.1)\times 10^{21}$~cm$^2$
  corresponding to $E(B-V)=0.63\pm0.02$. For $G<18$~mag and
  a distance of $2$~kpc, we can search for companions down
  to $>2M_\odot$.

\subsection{G263.9$-$03.3 (Vela)}

The Vela pulsar (PSR~J0835$-$4510, \citealt{Large1968}) has a well-measured
parallax ($\varpi = 3.5 \pm 0.2$) and proper motion ($-49.68\pm0.06$, $29.9\pm 0.1$)~mas/year
from \cite{Dodson2003}. 
We could 
find no explicit estimate for the center of the Vela SNR. \cite{Aschenbach1995}
clearly made such an estimate, as they report it is $25 \pm5$~arcmin from
the pulsar, but do not include the actual coordinates.  They report
the diameter of the remnant as $8.3$~degrees.  The position given by
\cite{Green2014} is visibly offset from the center of the X-ray image 
of the SNR in \cite{Aschenbach1995} and the reported diameter is only
$4.3$~degrees.  We set the center of the SNR to be $25$~arcmin backwards
from the neutron star along its proper motion vector, which should be the
\cite{Aschenbach1995} center and use their larger diameter.
\cite{Taylor1993} give a spin down age of 11 thousand years.
Estimates based on the properties of the SNR by 
\cite{Aschenbach1995} range from 8 to 37
thousand years.
$N(H) = (3.0\pm 0.3) \times 10^{20}$~cm$^2$ (\citealt{Pavlov2001}).
This column density implies $E(B-V) \simeq 0.05$~mag.
With its proximity and low extinction, there is no problem
searching for companions down to $>1M_\odot$. 

\subsection{G284.3$-$01.8}

  This SNR is associated with the $\gamma$-ray source
  1FGL~J1018.6$-$5856, and it was identified as an HMXB by \cite{Corbet2011}
  with the star 2MASS~J10185560$-$5856459.
  There is also an X-ray pulsar, PSR~J1016$-$5857, on the edge or just
  outside the remnant (\citealt{Camilo2001}).
  Like HESS~J0632$+$057, the high energy emission is believed to
  be due to interactions between a pulsar and the stellar wind
  (see the review by \citealt{Dubus2013}) and both \cite{Waisberg2015}
  and \cite{Strader2015} favor a neutron star as the compact companion.
  We adopt the position and size of the SNR from \cite{Green2019}.
  \cite{Ruiz1986} estimate an age of $10^4$~years.

\subsection{G296.5$+$10.0}

 G296.5$+$10.0 is associated with the X-ray PSR~J1210$-$5226 (\citealt{Zavlin2000}).
 \cite{Halpern2015} measure a proper motion of ($-12 \pm 5$, $9\pm 8$)~mas/year. 
 \cite{Giacani2000} estimate a distance of $2.1_{-0.8}^{+1.8}$~kpc, so we adopt
 a ``parallax'' of $\varpi = 0.50\pm 0.25$.  We adopt the center and diameter
 (76') from \cite{Green2019}. \cite{Roger1988} estimate an age of $7000$~year
 with uncertainties of a factor of $3$.
 \cite{deLuca2004} find $N(H)= (1.3\pm 0.1)\times 10^{21}$~cm$^2$, corresponding
 to $E(B-V) \simeq 0.22 \pm 0.02$, so for a distance of $4$~kpc we can probe
 down to $>2M_\odot$.

\section*{Data Availability Statement}

All data used in this paper is publicly available.


\begin{thebibliography}{} 

\bibitem[Abbott et al.(2021)]{Abbott2021} Abbott, R., Abbott, T.~D., Abraham, S., et al.\ 2021, \apjl, 913, L7
\bibitem[Anderson et al.(1996)]{Anderson1996} Anderson, S.~B., Cadwell, B.~J., Jacoby, B.~A., et al.\ 1996, \apjl, 468, L55
\bibitem[Aschenbach et al.(1995)]{Aschenbach1995} Aschenbach, B., Egger, R., \& Tr{\"u}mper, J.\ 1995, \nat, 373, 587 
\bibitem[\protect\citeauthoryear{Belczynski et al.}{2008}]{Belczynski2008} Belczynski, K., Kalogera, V., Rasio, F.~A., et al.\ 2008, \apjs, 174, 223
\bibitem[Benvenuto et al.(2013)]{Benvenuto2013} Benvenuto, O.~G., Bersten, M.~C., \& Nomoto, K.\ 2013, \apj, 762, 74
\bibitem[Bietenholz(2006)]{Bietenholz2006} Bietenholz, M.~F.\ 2006, \apj, 645, 1180 
\bibitem[Bietenholz et al.(2013)]{Bietenholz2013} Bietenholz, M.~F., Kondratiev, V., Ransom, S., et al.\ 2013, \mnras, 431, 2590
\bibitem[Blundell \& Bowler(2004)]{Blundell2004} Blundell, K.~M., \& Bowler, M.~G.\ 2004, \apjl, 616, L159
\bibitem[Boubert et al.(2017)]{Boubert2017} Boubert, D., Fraser, M., Evans, N.~W., Green, D.~A., \& Izzard, R.~G.\ 2017, \aap, 606, A14
\bibitem[Bovy et al.(2016)]{Bovy2016} Bovy, J., Rix, H.-W., Green, G.~M., et al.\ 2016, \apj, 818, 130
\bibitem[Camilo et al.(2001)]{Camilo2001} Camilo, F., Bell, J.~F., Manchester, R.~N., et al.\ 2001, \apjl, 557, L51
\bibitem[Camilo et al.(2002)]{Camilo2002} Camilo, F., Stairs, I.~H., Lorimer, D.~R., et al.\ 2002, \apjl, 571, L41
\bibitem[Castelli \& Kurucz(2003)]{Castelli2003} Castelli, F., \& Kurucz, R.~L.\ 2003, Modelling of Stellar Atmospheres, 210, 20P 
\bibitem[Cha et al.(1999)]{Cha1999} Cha, A.~N., Sembach, K.~R., \& Danks, A.~C.\ 1999, \apjl, 515, L25. 
\bibitem[Chatterjee et al.(2009)]{Chatterjee2009} Chatterjee, S., Brisken, W.~F., Vlemmings, W.~H.~T., et al.\ 2009, \apj, 698, 250 
\bibitem[Cigan et al.(2019)]{Cigan2019} Cigan, P., Matsuura, M., Gomez, H.~L., et al.\ 2019, \apj, 886, 51. 
\bibitem[Clark \& Stephenson(1977)]{Clark1977} Clark, D.~H., \& Stephenson, F.~R.\ 1977, Oxford [Eng.] ; New York : Pergamon Press, 1977.~1st ed.
\bibitem[Corbet et al.(2011)]{Corbet2011} Corbet, R.~H.~D., Cheung, C.~C., Kerr, M., et al.\ 2011, The Astronomer's Telegram, 3221,
\bibitem[Cutri et al.(2012)]{Cutri2012} Cutri, R.~M. \& et al.\ 2012, VizieR Online Data Catalog, II/311
\bibitem[de Luca et al.(2004)]{deLuca2004} de Luca, A., Mereghetti, S., Caraveo, P.~A., et al.\ 2004, \aap, 418, 625
\bibitem[Din{\c c}el et al.(2015)]{Dincel2015} Din{\c c}el, B., Neuh{\"a}user, R., Yerli, S.~K., et al.\ 2015, \mnras, 448, 3196
\bibitem[Dodson et al.(2003)]{Dodson2003} Dodson, R., Legge, D., Reynolds, J.~E., \& McCulloch, P.~M.\ 2003, \apj, 596, 1137 
\bibitem[Drimmel et al.(2003)]{Drimmel2003} Drimmel, R., Cabrera-Lavers, A., \& L{\'o}pez-Corredoira, M.\ 2003, \aap, 409, 205
\bibitem[Dubus(2013)]{Dubus2013} Dubus, G.\ 2013, A\&ARv, 21, 64
\bibitem[Eggleton(1983)]{Eggleton1983} Eggleton, P.~P.\ 1983, \apj, 268, 368
\bibitem[Eldridge et al.(2013)]{Eldridge2013} Eldridge, J.~J., Fraser, M., Smartt, S.~J., et al.\ 2013, \mnras, 436, 774
\bibitem[Eldridge et al.(2017)]{Eldridge2017} Eldridge, J.~J., Stanway, E.~R., Xiao, L., et al.\ 2017, \pasa,
\bibitem[Fraser \& Boubert(2019)]{Fraser2019} Fraser, M., \& Boubert, D.\ 2019, \apj, 871, 92 
\bibitem[Fruchter et al.(1988)]{Fruchter1988} Fruchter, A.~S., Taylor, J.~H., Backer, D.~C., Clifton, T.~R., \& Foster, R.~S.\ 1988, \nat, 331, 53 
\bibitem[Gaia Collaboration et al.(2016)]{Gaia2016} Gaia Collaboration, Prusti, T., de Bruijne, J.~H.~J., et al.\ 2016, \aap, 595, A1
\bibitem[Gaia Collaboration et al.(2021)]{Gaia2021} Gaia Collaboration, Brown, A.~G.~A., Vallenari, A., et al.\ 2021, \aap, 649, A1
\bibitem[Giacani et al.(2000)]{Giacani2000} Giacani, E.~B., Dubner, G.~M., Green, A.~J., et al.\ 2000, \aj, 119, 281
\bibitem[Gotthelf et al.(2007)]{Gotthelf2007} Gotthelf, E.~V., Helfand, D.~J., \& Newburgh, L.\ 2007, \apj, 654, 267 
\bibitem[Gotthelf \& Halpern(2009)]{Gotthelf2009} Gotthelf, E.~V., \& Halpern, J.~P.\ 2009, \apjl, 695, L35
\bibitem[Greco et al.(2021)]{Greco2021} Greco, E., Miceli, M., Orlando, S., et al.\ 2021, \apjl, 908, L45.
\bibitem[Green(2014)]{Green2014} Green, D.~A.\ 2014, Bulletin of the Astronomical Society of India, 42, 47
\bibitem[Green et al.(2019)]{Green2019b} Green, G.~M., Schlafly, E., Zucker, C., et al.\ 2019, \apj, 887, 93
\bibitem[Green(2019)]{Green2019} Green, D.~A.\ 2019, Journal of Astrophysics and Astronomy, 40, 36.
\bibitem[Gregory \& Fahlman(1980)]{Gregory1980} Gregory, P.~C., \& Fahlman, G.~G.\ 1980, \nat, 287, 805
\bibitem[Groh et al.(2013)]{Groh2013} Groh, J.~H., Meynet, G., Georgy, C., et al.\ 2013, \aap, 558, A131
\bibitem[Guseinov et al.(2005)]{Guseinov2005} Guseinov, O.~H., Ankay, A., \& Tagieva, S.~O.\ 2005, Astrophysics, 48, 330 
\bibitem[Gustafsson et al.(2008)]{Gustafsson2008} Gustafsson, B., Edvardsson, B., Eriksson, K., et al.\ 2008, \aap, 486, 951
\bibitem[Halpern \& Gotthelf(2015)]{Halpern2015} Halpern, J.~P. \& Gotthelf, E.~V.\ 2015, \apj, 812, 61.
\bibitem[Hamers et al.(2021)]{Hamers2021} Hamers, A.~S., Rantala, A., Neunteufel, P., et al.\ 2021, \mnras, 502, 4479
\bibitem[Han et al.(2020)]{Han2020} Han, Z.-W., Ge, H.-W., Chen, X.-F., et al.\ 2020, Research in Astronomy and Astrophysics, 20, 161
\bibitem[Henden et al.(2016)]{Henden2016} Henden, A.~A., Templeton, M., Terrell, D., et al.\ 201
\bibitem[Hillwig \& Gies(2008)]{Hillwig2008} Hillwig, T.~C., \& Gies, D.~R.\ 2008, \apjl, 676, L37
\bibitem[Hinton et al.(2009)]{Hinton2009} Hinton, J.~A., Skilton, J.~L., Funk, S., et al.\ 2009, \apjl, 690, L101
\bibitem[Hobbs et al.(2005)]{Hobbs2005} Hobbs, G., Lorimer, D.~R., Lyne, A.~G., \& Kramer, M.\ 2005, \mnras, 360, 974
\bibitem[Hoffleit \& Warren(1995)]{Hoffleit1995} Hoffleit, D. \& Warren, W.~H.\ 1995, VizieR Online Data Catalog, V/50x
\bibitem[Hohle et al.(2010)]{Hohle2010} Hohle, M.~M., Neuh{\"a}user, R., \& Schutz, B.~F.\ 2010, Astronomische Nachrichten, 331, 349
\bibitem[Hui \& Becker(2006)]{Hui2006} Hui, C.~Y., \& Becker, W.\ 2006, \aap, 454, 543
\bibitem[Israelian et al.(1999)]{Israelian1999} Israelian, G., Rebolo, R., Basri, G., et al.\ 1999, \nat, 401, 142.
\bibitem[Jayasinghe et al.(2021)]{Jayasinghe2021} Jayasinghe, T., Stanek, K.~Z., Thompson, T.~A., et al.\ 2021, \mnras, 504, 2577
\bibitem[Kerzendorf et al.(2019)]{Kerzendorf2019} Kerzendorf, W.~E., Do, T., de Mink, S.~E., et al.\ 2019, \aap, 623, A34 
\bibitem[\protect\citeauthoryear{Kobulnicky \& Fryer}{2007}]{Kobulnicky2007} Kobulnicky, H.~A., \& Fryer, C.~L.\ 2007, \apj, 670, 747
\bibitem[Kochanek(2014)]{Kochanek2014} Kochanek, C.~S.\ 2014, \apj, 785, 28.
\bibitem[Kochanek(2018)]{Kochanek2018} Kochanek, C.~S.\ 2018, \mnras, 473, 1633 
\bibitem[Kochanek et al.(2019)]{Kochanek2019} Kochanek, C.~S., Auchettl, K., \& Belczynski, K.\ 2019, \mnras, 485, 5394 
\bibitem[Koo et al.(1990)]{Koo1990} Koo, B.-C., Reach, W.~T., Heiles, C., Fesen, R.~A., \& Shull, J.~M.\ 1990, \apj, 364, 178 
\bibitem[Koo et al.(1993)]{Koo1993} Koo, B.-C., Yun, M.-S., Ho, P.~T.~P., \& Lee, Y.\ 1993, \apj, 417, 196
\bibitem[Kothes et al.(2006)]{Kothes2006} Kothes, R., Fedotov, K., Foster, T.~J., \& Uyan{\i}ker, B.\ 2006, \aap, 457, 1081 
\bibitem[Kothes \& Foster(2012)]{Kothes2012} Kothes, R., \& Foster, T.\ 2012, \apjl, 746, L4 
\bibitem[Kramer et al.(2003)]{Kramer2003} Kramer, M., Lyne, A.~G., Hobbs, G., et al.\ 2003, \apjl, 593, L31 
\bibitem[Krause et al.(2008)]{Krause2008} Krause, O., Birkmann, S.~M., Usuda, T., et al.\ 2008, Science, 320, 1195
\bibitem[Kulkarni et al.(1988)]{Kulkarni1988} Kulkarni, S.~R., Clifton, T.~C., Backer, D.~C., Foster, R.~S., \& Fruchter, A.~S.\ 1988, \nat, 331, 50
\bibitem[Lam et al.(2020)]{Lam2020} Lam, C.~Y., Lu, J.~R., Hosek, M.~W., et al.\ 2020, \apj, 889, 31
\bibitem[Large et al.(1968)]{Large1968} Large, M.~I., Vaughan, A.~E., \& Mills, B.~Y.\ 1968, \nat, 220, 340
\bibitem[Leahy \& Ranasinghe(2012)]{Leahy2012} Leahy, D.~A., \& Ranasinghe, S.\ 2012, \mnras, 423, 718
\bibitem[Lewandowski et al.(2004)]{Lewandowski2004} Lewandowski, W., Wolszczan, A., Feiler, G., Konacki, M., \& So{\l}tysi{\'n}ski, T.\ 2004, \apj, 600, 905 
\bibitem[Livingstone et al.(2009)]{Livingstone2009} Livingstone, M.~A., Ransom, S.~M., Camilo, F., et al.\ 2009, \apj, 706, 1163 
\bibitem[Lux et al.(2021)]{Lux2021} Lux, O., Neuh{\"a}user, R., Mugrauer, M., et al.\ 2021, Astronomische Nachrichten, 342, 553. 
\bibitem[Margon(1984)]{Margon1984} Margon, B.\ 1984, \araa, 22, 507
\bibitem[Marshall et al.(2006)]{Marshall2006} Marshall, D.~J., Robin, A.~C., Reyl{\'e}, C., et al.\ 2006, \aap, 453, 635
\bibitem[Marshall et al.(2013)]{Marshall2013} Marshall, H.~L., Canizares, C.~R., Hillwig, T., et al.\ 2013, \apj, 775, 75
\bibitem[Mayer \& Becker(2021)]{Mayer2021} Mayer, M.~G.~F. \& Becker, W.\ 2021, arXiv:2106.00700
\bibitem[Michalik et al.(2015)]{Michalik2015} Michalik, D., Lindegren, L., \& Hobbs, D.\ 2015, \aap, 574, A115
\bibitem[Moe \& Di Stefano(2017)]{Moe2016} Moe, M., \& Di Stefano, R.\ 2017, \apjs, 230, 15
\bibitem[Murray et al.(2002)]{Murray2002} Murray, S.~S., Slane, P.~O., Seward, F.~D., Ransom, S.~M., \& Gaensler, B.~M.\ 2002, \apj, 568, 226
\bibitem[Ng et al.(2007)]{Ng2007} Ng, C.-Y., Romani, R.~W., Brisken, W.~F., Chatterjee, S., \& Kramer, M.\ 2007, \apj, 654, 487
\bibitem[Ng \& Romani(2007)]{Ng2007b} Ng, C.-Y. \& Romani, R.~W.\ 2007, \apj, 660, 1357.
\bibitem[Ogata et al.(2021)]{Ogata2021} Ogata, M., Hirai, R., \& Hijikawa, K.\ 2021, \mnras, 505, 2485.
\bibitem[Orosz et al.(2001)]{Orosz2001} Orosz, J.~A., Kuulkers, E., van der Klis, M., et al.\ 2001, \apj, 555, 489.   
\bibitem[Panferov(2017)]{Panferov2017} Panferov, A.~A.\ 2017, \aap, 599, A77 
\bibitem[Pastorelli et al.(2020)]{Pastorelli2020} Pastorelli, G., Marigo, P., Girardi, L., et al.\ 2020, \mnras, 498, 3283
\bibitem[Patel et al.(2001)]{Patel2001} Patel, S.~K., Kouveliotou, C., Woods, P.~M., et al.\ 2001, \apjl, 563, L45
\bibitem[Pavlov et al.(2001)]{Pavlov2001} Pavlov, G.~G., Zavlin, V.~E., Sanwal, D., Burwitz, V., \& Garmire, G.~P.\ 2001a, \apjl, 552, L129
\bibitem[Pavlov et al.(2001b)]{Pavlov2001b} Pavlov, G.~G., Sanwal, D., K{\i}z{\i}ltan, B., \& Garmire, G.~P.\ 2001b, \apjl, 559, L131
\bibitem[Pejcha \& Thompson(2015)]{Pejcha2015} Pejcha, O. \& Thompson, T.~A.\ 2015, \apj, 801, 90
\bibitem[Perryman et al.(1997)]{Perryman1997} Perryman, M.~A.~C., Lindegren, L., Kovalevsky, J., et al.\ 1997, \aap, 500, 501
\bibitem[Podsiadlowski(1992)]{Podsiadlowski1992} Podsiadlowski, P.\ 1992, \pasp, 104, 717
\bibitem[Kameswara Rao et al.(2020)]{Rao2020} Kameswara Rao, N., Lambert, D.~L., Reddy, A.~B.~S., et al.\ 2020, \mnras, 493, 497.
\bibitem[Renzo et al.(2019)]{Renzo2019} Renzo, M., Zapartas, E., de Mink, S.~E., et al.\ 2019, \aap, 624, A66
\bibitem[Rest et al.(2011)]{Rest2011} Rest, A., Foley, R.~J., Sinnott, B., et al.\ 2011, \apj, 732, 3
\bibitem[Reynolds \& Aller(1988)]{Reynolds1988} Reynolds, S.~P., \& Aller, H.~D.\ 1988, \apj, 327, 845 
\bibitem[Reynoso et al.(2003)]{Reynoso2003} Reynoso, E.~M., Green, A.~J., Johnston, S., et al.\ 2003, \mnras, 345, 671 
\bibitem[Reynoso et al.(2017)]{Reynoso2017} Reynoso, E.~M., Cichowolski, S., \& Walsh, A.~J.\ 2017, \mnras, 464, 3029
\bibitem[Roberts et al.(1993)]{Roberts1993} Roberts, D.~A., Goss, W.~M., Kalberla, P.~M.~W., Herbstmeier, U., \& Schwarz, U.~J.\ 1993, \aap, 274, 427
\bibitem[Roger et al.(1988)]{Roger1988} Roger, R.~S., Milne, D.~K., Kesteven, M.~J., et al.\ 1988, \apj, 332, 940.
\bibitem[Ruiz \& May(1986)]{Ruiz1986} Ruiz, M.~T. \& May, J.\ 1986, \apj, 309, 667. 
\bibitem[Sana et al.(2012)]{Sana2012} Sana, H., de Mink, S.~E., de Koter, A., et al.\ 2012, Science, 337, 444
\bibitem[S{\'a}nchez-Cruces et al.(2018)]{Sanchez2018} S{\'a}nchez-Cruces, M., Rosado, M., Fuentes-Carrera, I., \& Ambrocio-Cruz, P.\ 2018, \mnras, 473, 1705 
\bibitem[Skrutskie et al.(2006)]{Skrutskie2006} Skrutskie, M.~F., Cutri, R.~M., Stiening, R., et al.\ 2006, \aj, 131, 1163
\bibitem[Slane et al.(2004)]{Slane2004} Slane, P., Helfand, D.~J., van der Swaluw, E., \& Murray, S.~S.\ 2004, \apj, 616, 403
\bibitem[Smith et al.(2011)]{Smith2011} Smith, N., Li, W., Filippenko, A.~V., et al.\ 2011, \mnras, 412, 1522
\bibitem[Sofue et al.(1980)]{Sofue1980} Sofue, Y., Furst, E., \& Hirth, W.\ 1980, \pasj, 32, 1 
\bibitem[Strader et al.(2015)]{Strader2015} Strader, J., Chomiuk, L., Cheung, C.~C., Salinas, R., \& Peacock, M.\ 2015, \apjl, 813, L26
\bibitem[Sukhbold et al.(2016)]{Sukhbold2016} Sukhbold, T., Ertl, T., Woosley, S.~E., et al.\ 2016, \apj, 821, 38
\bibitem[Taylor et al.(1993)]{Taylor1993} Taylor, J.~H., Manchester, R.~N., \& Lyne, A.~G.\ 1993, \apjs, 88, 529 
\bibitem[Tendulkar et al.(2013)]{Tendulkar2013} Tendulkar, S.~P., Cameron, P.~B., \& Kulkarni, S.~R.\ 2013, \apj, 772, 31
\bibitem[Thompson et al.(1978)]{Thompson1978} Thompson, G.~I., Nandy, K., Jamar, C., et al.\ 1978, 
\bibitem[Thompson et al.(2019)]{Thompson2019} Thompson, T.~A., Kochanek, C.~S., Stanek, K.~Z., et al.\ 2019, Science, 366, 637
\bibitem[Toonen et al.(2020)]{Toonen2020} Toonen, S., Portegies Zwart, S., Hamers, A.~S., et al.\ 2020, \aap, 640, A16
\bibitem[van den Bergh(1980)]{vandenBergh1980} van den Bergh, S.\ 1980, Journal of Astrophysics and Astronomy, 1, 67
\bibitem[van Leeuwen(2007)]{vanLeeuwen2007} van Leeuwen, F.\ 2007, \aap, 474, 653
\bibitem[Verbiest et al.(2012)]{Verbiest2012} Verbiest, J.~P.~W., Weisberg, J.~M., Chael, A.~A., Lee, K.~J., \& Lorimer, D.~R.\ 2012, \apj, 755, 39
\bibitem[Villase{\~n}or et al.(2021)]{Villasenor2021} Villase{\~n}or, J.~I., Taylor, W.~D., Evans, C.~J., et al.\ 2021, arXiv:2107.10170
\bibitem[Waisberg \& Romani(2015)]{Waisberg2015} Waisberg, I.~R., \& Romani, R.~W.\ 2015, \apj, 805, 18
\bibitem[Welsh et al.(2001)]{Welsh2001} Welsh, B.~Y., Sfeir, D.~M., Sallmen, S., \& Lallement, R.\ 2001, \aap, 372, 516 
\bibitem[Wenger et al.(2000)]{Wenger2000} Wenger, M., Ochsenbein, F., Egret, D., et al.\ 2000, A\&AS, 143, 9
\bibitem[Williamon et al.(2016)]{Williamon2016} Williamon, R.~M., Dale, H., Evavold, C.~L., et al.\ 2016, \pasp, 128, 124202.
\bibitem[Winkler et al.(1988)]{Winkler1988} Winkler, P.~F., Tuttle, J.~H., Kirshner, R.~P., \& Irwin, M.~J.\ 1988, IAU Colloq.~101: Supernova Remnants and the Interstellar Medium, 65 
\bibitem[Winkler et al.(2005)]{Winkler2005} Winkler, P.~F., Long, K.~S., Hamilton, A.~J.~S., et al.\ 2005, \apj, 624, 189
\bibitem[Yao et al.(2021)]{Yao2021} Yao, J., Zhu, W., Manchester, R.~N., et al.\ 2021, Nature Astronomy
\bibitem[Zacharias et al.(2005)]{Zacharias2005} Zacharias, N., Monet, D.~G., Levine, S.~E., et al.\ 2005, VizieR Online Data Catalog, I/297
\bibitem[Zavlin et al.(2000)]{Zavlin2000} Zavlin, V.~E., Pavlov, G.~G., Sanwal, D., \& Tr{\"u}mper, J.\ 2000, \apjl, 540, L25
\bibitem[Zeiger et al.(2008)]{Zeiger2008} Zeiger, B.~R., Brisken, W.~F., Chatterjee, S., \& Goss, W.~M.\ 2008, \apj, 674, 271 
\end{thebibliography}
\end{document}